\documentclass[aps,prb,superscriptaddress,twocolumn,floatfix]{revtex4-1}

\usepackage{amsmath}
\usepackage{amssymb}
\usepackage{amsthm}
\usepackage{graphicx}
\usepackage{graphics}
\usepackage{subfigure}
\usepackage{color}
\usepackage{bm}
\usepackage{hyperref}
\usepackage{rotating}
\def\t#1{\widetilde{#1}}
\def\wb#1{\overline{#1}}

\newcommand{\<}{\langle}
\renewcommand{\>}{\rangle}
\newcommand{\be}{\begin{equation} }
\newcommand{\ee}{\end{equation} }
\newcommand{\ba}{\begin{eqnarray} }
\newcommand{\ea}{\end{eqnarray} }

\newcommand{\down}{\downarrow}

\newcommand{\bpm}{\begin{pmatrix}}
\newcommand{\epm}{\end{pmatrix}}
\newcommand{\bmm}{\begin{matrix}}
\newcommand{\emm}{\end{matrix}}
\newcommand{\up}{\uparrow}

\begin{document}

\title{Braiding statistics approach to symmetry-protected topological phases}

\author{Michael Levin}
\affiliation{Condensed Matter Theory Center, Department of Physics, University of Maryland, College Park, Maryland 20742, USA}

\author{Zheng-Cheng Gu}
\affiliation{Kavli Institute for Theoretical Physics, University of
California, Santa Barbara, CA 93106, USA}

\date{\today}

\begin{abstract}
We construct a 2D quantum spin model that realizes an Ising paramagnet with gapless edge modes protected by
Ising symmetry. This model provides an example of a ``symmetry-protected topological phase.''
We describe a simple physical construction that distinguishes this system from a conventional paramagnet:
we couple the system to a $\mathbb{Z}_2$ gauge field and then show that the $\pi$-flux excitations have different braiding
statistics from that of a usual paramagnet. In addition, we show that these braiding statistics directly imply
the existence of protected edge modes. Finally, we analyze a particular microscopic model for the edge and derive 
a field theoretic description of the low energy excitations. We believe that the braiding statistics approach outlined
in this paper can be generalized to a large class of symmetry-protected topological phases.
\end{abstract}


\maketitle

\section{Introduction}
We now know that there are two distinct types of
time reversal invariant band insulators: topological insulators
and conventional insulators.\cite{KaneMele,KaneMele2, Roy,FuKaneMele,MooreBalents,HasanKaneRMP}
The two families of insulators are distinguished by the fact that topological insulators have
protected gapless boundary modes while trivial insulators do not. 
It is important to remember that time reversal and charge conservation symmetry play
a crucial role in this physics: if either of these symmetries are broken 
(either explicitly or spontaneously), the boundary modes
can be gapped out and the sharp distinction between topological insulators and
conventional insulators disappears.

This observation motivates a generalization of topological insulators called
``symmetry-protected topological (SPT) phases''\cite{GuSPT, XieSPT1, XieSPT2,XieSPT3,XieSPT4,PollmannSPT1,PollmannSPT2,NorbertSPT,LukaszfSPT1}.
To define this concept, consider a general quantum many-body system. The system may be built out of fermions or bosons/spins,
and can live in any spatial dimension. We will say that such a system belongs to a nontrivial SPT phase if it satisfies 
four properties. The first property is that the system has a finite energy gap to excitations in the bulk. The second property is 
that the Hamiltonian is invariant under some set of internal (on-site) symmetries, and none
of these symmetries are broken spontaneously. The third property is that the ground state belongs to a distinct quantum phase 
from a ``trivial state'' with the same symmetry. That is, one cannot continuously connect the ground state 
with a ``trivial state'' without breaking one of the symmetries or closing the energy gap. Here, by
a ``trivial state'', we mean a product state (in the boson/spin case) or an atomic insulator (in the fermion case). The final
property of an SPT phase is that the ground state \emph{can} be continuously connected with a
trivial state without closing the energy gap if one or more of the symmetries are broken during the process.
We note that nontrivial SPT phases typically exhibit robust gapless boundary modes analogous to that
of topological insulators, though we will not include this property in the formal definition.

Symmetry-protected topological phases have a long history
in the one dimensional (1D) case. Most famously, the Haldane phase of the $S=1$ Heisenberg
antiferromagnet\cite{Haldanephase} is known to belong to this class\cite{GuSPT,PollmannSPT1,PollmannSPT2}.
More recently, a complete classification of 1D SPT phases was obtained for both
boson/spin systems\cite{XieSPT1,NorbertSPT,XieSPT2} and fermion systems.\cite{LukaszfSPT1,XieSPT2}

Much less is known about higher dimensional SPT phases. In the case of fermion systems, our understanding
is largely limited to non-interacting models such as topological insulators or superconductors. For these systems,
an (almost) complete classification of SPT phases was obtained by Ref. \onlinecite{RyuSPT,Kitaevperiod}. In some cases,
it is known that this classification scheme is not affected by interactions (e.g. the $\mathbb{Z}_2$ classification
of topological insulators in two\cite{FuKanepump} and three\cite{FuKaneHall,LevinBurnellKoch} dimensions).
In general, however, this need not be the case\cite{LukaszfSPT2} and consequently our understanding of interacting fermionic SPT phases
in higher dimensions is incomplete.

The boson case has received even less attention, and will be our focus here.
In this case, a major advance was made by the recent paper, Ref. \onlinecite{XieSPT4}. In that paper,
the authors proposed a general classification scheme for bosonic SPT phases in general spatial dimension.
Also, the authors constructed concrete microscopic models realizing each of these phases. This work established that
the boson case is tractable even for interacting systems.

Nevertheless, a number of questions remain open.
One problem is that we have not identified any physical properties that distinguish different SPT phases
in the bulk. The boundary physics is also poorly understood: while Ref. \onlinecite{XieSPT3} showed 
that the 2D SPT states have symmetry-protected gapless boundary modes, the problem for higher dimensions remains open.

In this work, we address these (and other) questions in the context of a simple example.
Specifically, we consider the case of 2D spin systems with a $\mathbb{Z}_2$ Ising-like symmetry. According to
Refs. \onlinecite{XieSPT3,XieSPT4}, there is exactly one nontrivial SPT phase with this symmetry. This phase
can be thought of as a new kind of Ising paramagnet. Here, we construct an exactly soluble spin model that realizes this phase.
We then derive three main results. Our first result is a simple argument that this model belongs to a distinct phase from a
conventional Ising paramagnet. We derive this result by coupling the model to a $\mathbb{Z}_2$ gauge field. After following this
procedure, we find that the resulting gauged spin model supports quasiparticle excitations with different braiding statistics
from that of a conventional (gauged) paramagnet. More specifically, we find that in a conventional paramagnet, the $\pi$-flux
excitations have bosonic or fermionic statistics, while in the new paramagnet they have semionic statistics.
It then follows immediately that the two paramagnets cannot be continuously connected without breaking the $\mathbb{Z}_2$ symmetry or
closing the energy gap. Closely related to this observation, we show that the two spin models are ``dual'' to two previously studied
lattice models -- each of which realizes a different type of $\mathbb{Z}_2$ gauge theory. This duality establishes a connection 
between SPT phases and previous work\cite{Wittencohomology} on the classification of topological gauge theories.

Our second result is a proof that the new paramagnet has gapless edge modes protected by Ising symmetry. Interestingly, our argument
reveals that the protected edge states are deeply connected to the braiding statistics of the $\pi$-fluxes.
This approach to proving edge state protection is somewhat different from the original argument of Ref. \onlinecite{XieSPT3}
and may be more amenable to higher dimensional generalizations. In the final part of the paper, we analyze the protected
edge modes at a more concrete level, focusing on a particular microscopic model of the edge. We derive a field theoretic
description of the low energy modes, and analyze their stability to perturbations.

Although we focus our discussion on a particular SPT phase, we believe that our basic approach is more general.
That is, we expect that in a large class of SPT phases, braiding statistics can be used to uniquely characterize
the bulk and to derive the existence of protected boundary modes.
We discuss these potential generalizations in the conclusion.

This paper is organized as follows. In section \ref{modelsect}, we describe spin models that realize both the
conventional and the new kind of Ising paramagnet. In section \ref{pifluxsect} we show that the two spin
models can be distinguished by the braiding statistics of the $\pi$-flux excitations. In section 
\ref{dualsect} we show that the two spin models are dual to two previously studied lattice models.
In section \ref{genargsect}, we show that the $\pi$-flux braiding statistics are directly connected to
the existence of protected edge modes. Finally, in section \ref{edgesect} we analyze a particular microscopic model for the edge.

\section{Two kinds of Ising paramagnets} \label{modelsect}
\begin{figure}[t]
{\includegraphics[width=0.95\columnwidth]{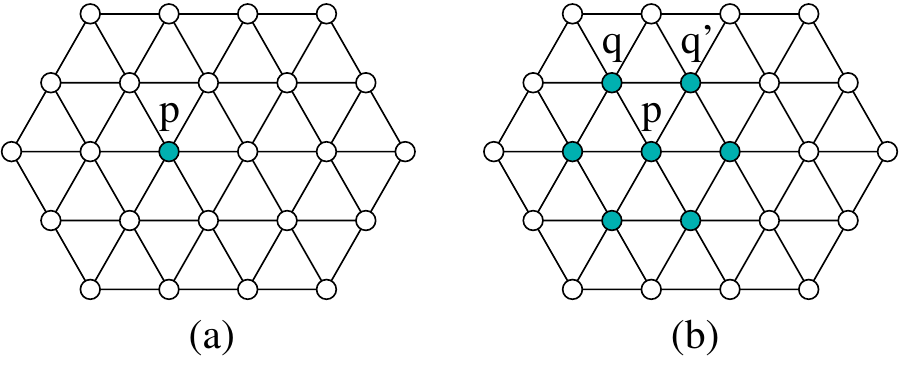}}
\caption{
The Hamiltonians $H_0, H_1$ (\ref{Hconv}-\ref{Hnew}) for the two spin models. (a) The Hamiltonian $H_0$ 
is a sum of single spin terms, $\sigma^x_p$. (b) The Hamiltonian $H_1$ is a sum of seven spin terms
$B_p = -\sigma^x_p \prod_{\<pqq'\>} i^{\frac{1-\sigma^z_{q}\sigma^z_{q'}}{2}}$ where the product runs over the
six triangles $\<pqq'\>$ containing $p$.
}\label{fig:Hspin}
\end{figure}

To begin, consider the following spin-$1/2$ model defined on the triangular lattice (Fig. \ref{fig:Hspin}a):
\begin{equation}
H_0 = - \sum_p \sigma^x_p
\label{Hconv}
\end{equation}
This model describes a (conventional) Ising paramagnet. To see this, note that the system
satisfies two properties. First, the Hamiltonian is invariant under the Ising symmetry
$S = \prod_p \sigma^x_p$. Second, the ground state $|\Psi_0\> \equiv |\sigma^x_p = 1\>$
is gapped and unique -- implying that the symmetry is not broken spontaneously.

Surprisingly, there is another type of Ising paramagnet which is qualitatively different
from $H_0$ and represents a distinct quantum phase. A microscopic model for this new type of paramagnet
was first constructed in Ref. \onlinecite{XieSPT3}. Here we describe another model which is
more convenient for our purposes. The model we consider is a spin-$1/2$ system on the triangular lattice.
The Hamiltonian is given by (Fig. \ref{fig:Hspin}b):
\begin{align}
H_1 = -\sum_p B_p \ , \ \ \ \
B_p = -\sigma^x_p \prod_{\<pqq'\>} i^{\frac{1-\sigma^z_{q}\sigma^z_{q'}}{2}}\label{Hnew}
\end{align}
where the product runs over the six triangles $\<pqq'\>$ containing the site $p$. We note that this
Hamiltonian is Hermitian despite the factors of $i$. To see this, notice that
the product includes a factor of $i$ for each pair of neighboring spins $q, q'$ that have opposite
values of $\sigma^z$. In particular, since the number of such pairs is necessarily even, the product always
reduces to a factor of $\pm 1$. It is then clear that $H_1^\dagger = H_1$.
(For readers who are curious as to how this model was constructed, see section \ref{dualsect}).

First we show that $H_1$ describes a paramagnetic phase -- that is, the Ising symmetry is not
spontaneously broken. To establish this fact, we solve $H_1$
explicitly. The key point is that 
\begin{equation}
[B_p, B_{p'}] = 0
\end{equation}
as can be verified by straightforward algebra. As a result, we can simultaneously diagonalize $\{B_p\}$.
We will label the simultaneous eigenstates by $|\{b_p\}\>$ where
$b_p = \pm 1$ denotes the eigenvalues of $B_p$. It is not hard to show that there
is a unique state for each choice of $\{b_p\}$, assuming a periodic geometry (i.e. a torus). In other words, 
the $\{b_p\}$ are a complete set of quantum numbers. We therefore have the full energy spectrum: each state 
$|b_p\>$ is an energy eigenstate with energy 
\begin{equation}
E = - \sum_p b_p
\end{equation}
In particular, the ground state $|\Psi_1\> \equiv |b_p = 1\>$ is unique and gapped --
implying that the Ising symmetry is not spontaneously broken.

\begin{figure}[t]
{\includegraphics[width=0.9\columnwidth]{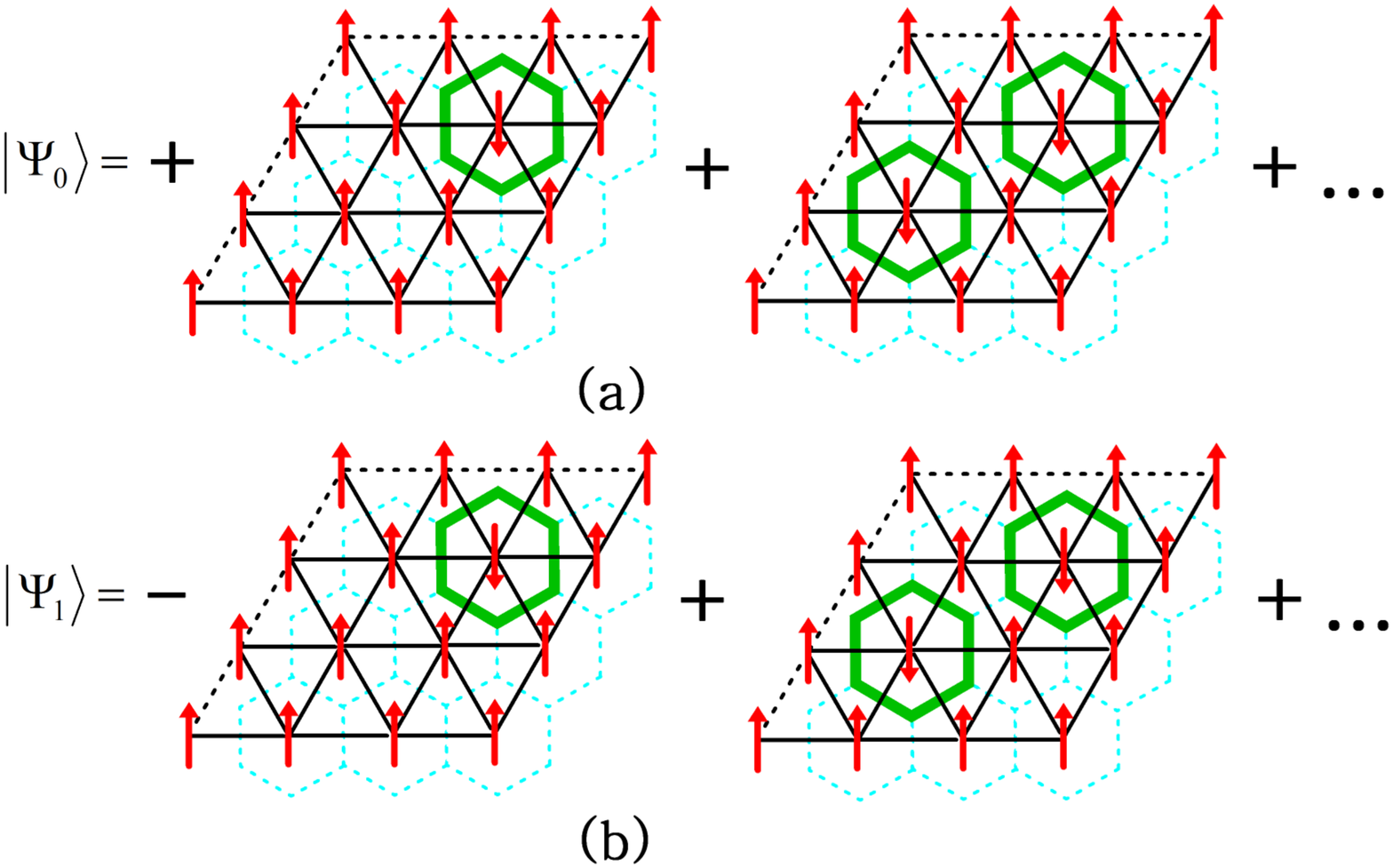}}
\caption{
A schematic plot of the ground states $\Psi_0$ and $\Psi_1$ for the two paramagnets
$H_0, H_1$. (a) In terms of domain wall configurations, the ground state $\Psi_0$ is a
equal weight superposition of all configurations. (b) The ground state $\Psi_1$ is also a
superposition of all domain wall configurations, but each configuration enters with a sign
$(-1)^{N_{dw}}$ where $N_{dw}$ is the total number of domain walls.
}\label{fig:wavefunction}
\end{figure}

It is illuminating to compare the ground state wave functions of $H_0, H_1$. The ground state
of $H_0$ is the state where $\sigma^x_p = 1$ everywhere. Working in the $\sigma^z$ basis,
the wave function is given by
\begin{equation}
\Psi_0(\{\alpha_p\}) = 1
\end{equation}
for all spin configurations $\{\alpha_p = \uparrow, \downarrow\}$ (Fig. \ref{fig:wavefunction}a).
As for $H_1$, we note that the ground state is the unique state with $B_p = 1$ everywhere. It is
straightforward to check that the corresponding wave function is given by
\begin{equation}
\Psi_1(\{\alpha_p\}) = (-1)^{N_{dw}}
\end{equation}
where $N_{dw}$ is the total number of domain walls in the
spin-configuration $\{\alpha_p = \uparrow, \downarrow\}$ (Fig. \ref{fig:wavefunction}b). We can see that the two ground states are
nearly identical, differing only by some phase factors. Nevertheless, these two states belong to
two different quantum phases, as we now show.

\section{Coupling the spin models to a \texorpdfstring{$Z_2$}{Z2} gauge field} \label{pifluxsect}
In this section, we show that $H_0, H_1$ belong to distinct quantum phases.
Our strategy is as follows. Because $H_0, H_1$ have a $\mathbb{Z}_2$ symmetry, we can couple them to a $\mathbb{Z}_2$ gauge field
$\mu^z_{pq} = \pm 1$ which lives on the links $\<pq\>$ of the triangular lattice. We then show that
the resulting gauged spin models have quasiparticle excitations with different braiding statistics.
More specifically, we show that the two systems differ in the statistics of the $\pi$-flux excitations: while 
the $\pi$-fluxes have bosonic or fermionic statistics in the case of $H_0$, they have semionic statistics 
in the case of $H_1$. It then follows immediately that $H_0, H_1$ cannot be continuously connected without 
breaking the $\mathbb{Z}_2$ symmetry or closing the energy gap.

\begin{figure}[t]
{\includegraphics[width=0.95\columnwidth]{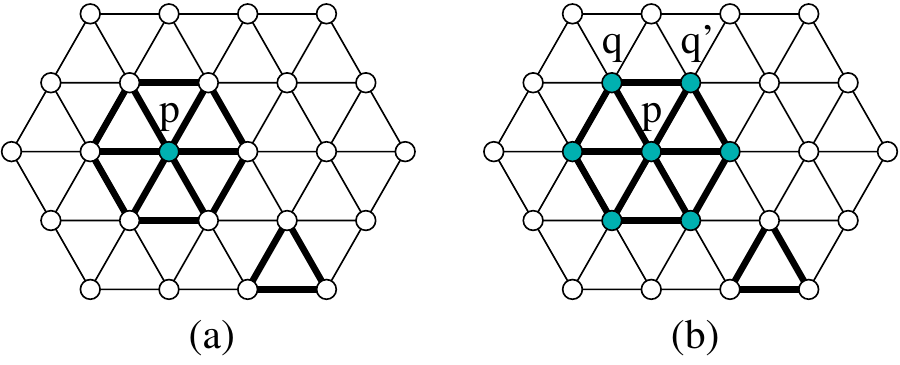}}
\caption{
The Hamiltonians $\t{H}_0,\t{H}_1$ (\ref{gaugeH}) for the two gauged spin models. (a) The Hamiltonian $\t{H}_0$
is a sum of two terms. The first term is the gauge flux term $\mu^z_{pq}\mu^z_{qr}\mu^z_{rp}$ (thick triangle) where $\mu^z_{pq}$ denotes the
$\mathbb{Z}_2$ gauge field on the link $\<pq\>$. The second term is the spin interaction $\sigma^x_p O_p$ where
$O_p = \prod_{\<pqr\>} (1 + \mu^z_{pq}\mu^z_{qr}\mu^z_{rp})/2$ and the product runs over the six triangles adjacent to $p$.
(b) The Hamiltonian $\t{H}_1$ includes the same gauge flux term $\mu^z_{pq}\mu^z_{qr}\mu^z_{rp}$ but has a more complicated
seven spin interaction $\t{B}_p O_p$ (\ref{Bgauge}).
}\label{fig:Hgaugespin}
\end{figure}

Coupling $H_0, H_1$ to a $\mathbb{Z}_2$ gauge field requires several steps.\cite{Kogut} The first step is to apply the minimal coupling
procedure, replacing nearest neighbor spin-spin interactions like $\sigma^z_{q} \sigma^z_{q'}$ with
$\sigma^z_{q} \mu^z_{qq'} \sigma^z_{q'}$. Next, we multiply each term in the resulting Hamiltonian (either $\sigma^x_p$ or $B_p$) 
by the operator
\begin{equation}
O_p = \prod_{\<pqr\>} (1 + \mu^z_{pq}\mu^z_{qr}\mu^z_{rp})/2
\end{equation}
where the product runs over the six triangles $\<pqr\>$ adjacent to site $p$. The operator $O_p$ is a projector which
projects onto states that have vanishing flux through each of the adjoining triangles. We include this projection 
operator in order to ensure that our gauged Hamiltonian is Hermitian, and also to make the minimal coupling procedure 
unambiguous. (For more general models, we would replace $O_p$ with an operator that projects onto states that have
vanishing flux through all the triangles in the vicinity of the spin-spin interactions). The final step is to add a
term of the form $- \sum_{\<pqr\>} \mu^z_{pq}\mu^z_{qr}\mu^z_{rp}$ to the Hamiltonian. This term ensures that
the states with vanishing $\mathbb{Z}_2$ flux have the lowest energy. The resulting models are given by (Fig. \ref{fig:Hgaugespin}):
\begin{eqnarray}
\t{H}_0 &=& - \sum_p \sigma^x_p O_p - \sum_{\<pqr\>} \mu^z_{pq}\mu^z_{qr}\mu^z_{rp} \nonumber \\
\t{H}_1 &=& - \sum_p \t{B}_p O_p - \sum_{\<pqr\>} \mu^z_{pq}\mu^z_{qr}\mu^z_{rp}
\label{gaugeH}
\end{eqnarray}
where
\begin{equation}
\t{B}_p = -\sigma^x_p \prod_{\<pqq'\>} i^{\frac{1-\sigma^z_{q}\mu^z_{qq'} \sigma^z_{q'}}{2}}
\label{Bgauge}
\end{equation}
Like all gauge theories, these models are defined on a Hilbert space consisting of gauge invariant states -- that is, 
all states satisfying the constraint
\begin{equation}
\prod_{q} \mu^x_{pq} = \sigma^x_p
\label{gauss}
\end{equation}
for all sites $p$.\cite{Kogut} This constraint can be thought of as a $\mathbb{Z}_2$ analog of Gauss' law, $\nabla \cdot E = 4\pi\rho$.

Importantly, all the terms in $\t{H}_0, \t{H}_1$ commute with one another so these Hamiltonians can be
solved exactly just like the ungauged spin models $H_0, H_1$. In particular, it is
easy to verify that both models have a finite energy gap. 

The next task is to construct the quasiparticle excitations and show 
that they have different braiding statistics in the two systems.
The quickest way to derive this fact is to note that $\t{H}_0, \t{H}_1$ can be exactly mapped onto
the previously studied ``toric code''\cite{KitaevToric, Stringnet} and ``doubled semion''\cite{Stringnet} models. These two models
have been analyzed in detail and are known to support quasiparticle excitations with different statistics.\cite{Stringnet}
A description of these models as well as the mapping to $\t{H}_0, \t{H}_1$ is given in section \ref{dualsect}.

Alternatively, we can directly compute the quasiparticle statistics of
$\t{H}_0, \t{H}_1$ and show that they are different. 
The first type of excitation is a ``spin-flip'', which we will denote by $e$. These
excitations correspond to sites $p$ where $\sigma^x_p = -1$ for the case of $\t{H}_0$, or $\t{B}_p = -1$
for the case of $\t{H}_1$. The second
type of excitation is the ``$\pi$-flux'', $m$. These excitations correspond to triangular plaquettes $\<pqr\>$
where $\mu^z_{pq} \mu^z_{qr} \mu^z_{rp} = -1$. In fact, there are \emph{two} types of
$\pi$-flux excitations, which differ by the addition of a spin-flip: $m_b = m_a \cdot e$.

It is clear that in both systems, if we braid a spin-flip excitation $e$ around either of the $\pi$-flux excitations
$m_a, m_b$, the resulting statistical Berry phase is $\pi$ (in some sense this is the definition of a $\pi$-flux excitation).
It is also intuitively clear that the spin-flip excitation $e$ is a boson in both models. All that remains
is to understand the statistics of the $\pi$-fluxes. As we will now show, this is where the two
models differ. 

\begin{figure}[t]
{\includegraphics[width=0.5\columnwidth]{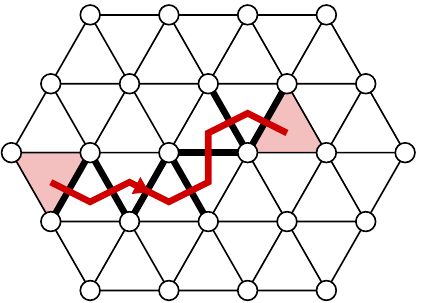}}
\caption{
The string operator $V^0_{\beta}$ (\ref{string0}) is defined for any path $\beta$ on the dual honeycomb lattice and is given by
a product of $\mu^x_{pq}$ over all links $\<pq\>$ crossing $\beta$ (thickened lines). Applying this operator to the ground state
$|\Psi_0\>$ creates two $\pi$-fluxes at the endpoints of $\beta$ (shaded triangles).}\label{fig:string0}
\end{figure}

To determine the $\pi$-flux statistics, we first identify operators that create these excitations. Like all
quasiparticles with nontrivial braiding statistics, the $\pi$-fluxes can be
created using an extended \emph{string-like} operator.\cite{LevinWenHop} If we apply these string-like operators to the ground
state, the result is a \emph{pair} of $\pi$-flux excitations -- one at each end of the string. In the case of
$\t{H}_0$, the following string operator does the job:
\begin{equation}
V^0_\beta = \prod_{\<pq\> \perp \beta} \mu^x_{pq}
\label{string0}
\end{equation}
Here $\beta$ is a path in the dual honeycomb lattice joining the two triangular plaquettes,
and the product runs over all links $\<pq\>$ crossing $\beta$ (Fig. \ref{fig:string0}). We can verify that $V^0_\beta$
creates $\pi$-flux excitations at the two endpoints of $\beta$ by noting that $V^0_\beta$ anticommutes
with the flux $\mu^z_{pq} \mu^z_{qr} \mu^z_{rp}$ through the two triangles at the ends of $\beta$. At the same
time, this operator commutes with all the other terms in $\t{H}_0$ so it does not create any additional 
excitations.\footnote{To be precise, the $\sigma^x_p O_p$ terms at the sites neighboring the endpoints of 
$\beta$ do not commute with $V^0_\beta$, but the excitations associated with these terms should be regarded as part of the $\pi$-flux 
excitation.} Closely related to this fact, one can check that the state $V^0_\beta |\Psi_0\>$ does not depend on the choice of path $\beta$, but only on the endpoints of $\beta$ -- a general feature of such
string-like operators.\cite{KitaevToric,LevinWenHop,Stringnet} We will denote the $\pi$-flux excitation created by $V^0_\beta$ by $m_a$.
A similar string operator creates the other type of $\pi$-flux, $m_b = m_a \cdot e$.

\begin{figure}[t]
{\includegraphics[width=0.7\columnwidth]{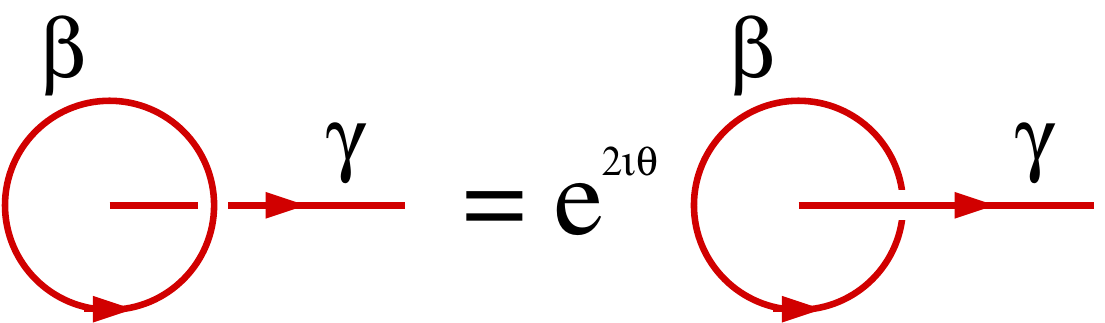}}
\caption{
A schematic picture of the two states $V^0_\beta V^0_\gamma |\Psi_0\>$, $V^0_\gamma V^0_\beta |\Psi_0\>$. The first state (left) is obtained from
a process in which two $\pi$-fluxes are created at the endpoints of $\gamma$, and then two more fluxes are created, braided around
the path $\beta$ and then annihilated. The second state (right) corresponds to executing these two steps in the opposite order.
We expect these two states to differ by the Berry phase $e^{2i\theta}$ associated
with braiding one $\pi$-flux excitation around another. The same is true for $V^1_\beta, V^1_\gamma$.
}\label{fig:statstr}
\end{figure}

In general, one of the most important aspects of string operators is the commutation relations satisfied by
two intersecting strings. Let $\beta, \gamma$ be two paths on the dual honeycomb lattice that intersect one
another. Using the definition (\ref{string0}), we can see that the two corresponding string operators
commute with one another:
\begin{equation}
V^0_{\beta} V^0_{\gamma} = V^0_{\gamma} V^0_{\beta}
\label{string0com}
\end{equation}
This string algebra is important because we can use it to find the statistics
of the quasiparticle $m_a$.\cite{KitaevToric,LevinWenHop,Stringnet}
One way to see this is to consider the special case where $\beta$ is a
\emph{closed} path and $\gamma$ is an open path, as in Fig. \ref{fig:statstr}. In this case, the two operators
$V^0_\beta$ and $V^0_\gamma$ have different physical interpretations: while the operator
$V^0_\gamma$ can be thought of describing a physical process in which two $\pi$-fluxes are created
and then moved to the endpoints of $\gamma$, the operator $V^0_\beta$ does not create any excitations at all.
In fact, it is easy to check that $V^0_\beta$ exactly commutes with the Hamiltonian $\t{H}_0$ whenever
$\beta$ forms a closed loop. This suggests that $V^0_\beta$ should be thought of as describing a three step process in
which (1) two $\pi$-fluxes are created, (2) one of the $\pi$-fluxes moves
all the way around the closed path $\beta$, and then (3) the two $\pi$-fluxes are annihilated. Using
this interpretation, we can see that the state $V^0_{\beta} V^0_{\gamma} |\Psi\>$ is the end result of
a process in which two $\pi$-fluxes are created at the endpoints of $\gamma$, and then afterwards
another $\pi$-flux is braided around one of the endpoints and annihilated with its partner.
In contrast, the state $V^0_{\gamma} V^0_\beta |\Psi\>$ corresponds to executing these two steps in
the opposite order. Comparing these two processes, we expect that they will differ by a phase factor which is exactly the
statistical Berry phase associated with braiding one $\pi$-flux around another. In other words,
the phase difference between these two states should be $e^{2i\theta}$ where $\theta$
is the exchange statistics for the particles:
\begin{equation}
V^0_{\beta} V^0_{\gamma} |\Psi_0\> = e^{2i\theta} \cdot V^0_{\gamma} V^0_{\beta} |\Psi_0\>
\label{stat}
\end{equation}
In light of this relation, equation (\ref{string0com}) implies that $\theta = 0$ or $\pi$.
That is, $m_a$ is either a boson or a fermion. A similar analysis 
shows that the other $\pi$-flux excitation, $m_b$, is also either a boson or fermion. In fact, with a bit more work
one can establish the more precise result that $m_a$ is boson and $m_b$ is a fermion. The difference in statistics between
$m_a, m_b$ comes from the fact that $m_b = m_a \cdot e$ where $e, m_a$ have mutual statistics $\pi$. However,
we will not need this more detailed result here. (See Refs. \onlinecite{Stringnet,LevinWenHop} for an analogous calculation 
for the closely related ``toric code'' model).

\begin{figure}[t]
{\includegraphics[width=0.5\columnwidth]{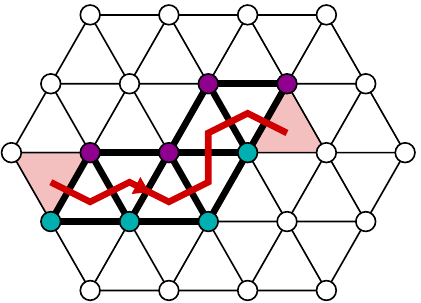}}
\caption{
(Color online) The string operator $V^1_\beta$ (\ref{string1}) is defined for any path $\beta$ on the dual honeycomb lattice. It
acts on all triangles $\<pqq'\>$ along the path $\beta$ (thickened lines). The action is different depending on whether $q,q'$
are to the left of $\beta$ (purple sites) or to the right of $\beta$ (blue sites). Applying this operator to the ground state $|\Psi_1\>$
creates two $\pi$-fluxes at the endpoints of $\beta$ (shaded triangles). }\label{fig:string1}
\end{figure}

We can repeat the same analysis for $\t{H}_1$. In this case, the following string operator creates a $\pi$-flux
excitation:
\begin{eqnarray}
V^1_{\beta} &=& \prod_{\<pq\> \perp \beta} \mu^x_{pq} \cdot \prod_{\<pqq'\>, r}
i^{\frac{1-\sigma^z_{q}\mu^z_{qq'}\sigma^z_{q'}}{2}}  \label{string1} \\
&\cdot& \prod_{\<pqq'\>,l} (-1)^{\t{s}_{pqq'}} \cdot \prod_{\<pqq'\> \in \beta} (1+ \mu^z_{pq}\mu^z_{qq'}\mu^z_{pq'})/2
\nonumber
\end{eqnarray}
Here, the first product runs over all links $\<pq\>$ crossing $\beta$. The next two products run over all triangles 
$\<pqq'\>$ along the path such that $q,q'$ are to the right of $\beta$
or to the left of $\beta$ respectively (Fig. \ref{fig:string1}). The last product runs over all triangles along
$\beta$. The operator $\t{s}_{pqq'}$ is defined by
\begin{equation}
\t{s}_{pqq'} = \frac{1}{4} (1-\sigma^z_p \mu^z_{pq} \sigma^z_q)(1+\sigma^z_p \mu^z_{pq'} \sigma^z_{q'})
\end{equation}
As in the previous case, one can check $V^1_\beta$ anticommutes with the flux $\mu^z_{pq} \mu^z_{qr} \mu^z_{rp}$
through the two triangles at the ends of $\beta$, but commutes with the Hamiltonian $\t{H}_1$ everywhere else. Hence, if
we apply $V^1_\beta$ to the ground state, it creates $\pi$-fluxes at the two endpoints of $\beta$. We will again denote
this $\pi$-flux excitation by $m_a$. (For readers who are curious, $V^1_\beta$ was constructed from the ``doubled semion model''
string operators\cite{Stringnet} using the exact mapping of section \ref{dualsect}).

In this case, one can check that the string operators satisfy a slightly different algebra: for any
two paths $\beta, \gamma$ intersecting one another, we have
\begin{equation}
V^1_{\beta} V^1_{\gamma} = -V^1_{\gamma} V^1_{\beta}
\label{string1com}
\end{equation}
Therefore by the same reasoning as in (\ref{stat}), we conclude that the statistical angle $\theta$ satisfies
$2\theta = \pi$, so that $\theta = \pm \pi/2$.
In other words, $m_a$ is a semion. A similar analysis shows that the other
$\pi$-flux excitation $m_b$ is also a semion. With a bit more work\cite{Stringnet}, one can show that $m_a, m_b$ have opposite
statistics -- that is $\theta = \pi/2$ in one case and $\theta = -\pi/2$ in the other -- but again we do not need this
more detailed result here.

We have shown that the $\pi$-fluxes have different statistics in the two gauged spin models: these
excitations are bosons or fermions in the case of $\t{H}_0$, and are semions in the case of $\t{H}_1$.
This result provides a simple physical distinction between the two systems. It also proves that the
two spin models $H_0, H_1$ cannot be continuously connected with one another without breaking the $\mathbb{Z}_2$ symmetry or
closing the energy gap. Indeed, if such
a path existed, then we could construct a corresponding path connecting the gauged spin models $\t{H}_0, \t{H}_1$
-- a contradiction. We note, however, that the above argument does not rule out the possibility of connecting
$H_0, H_1$ if the Ising symmetry is broken during the process. Indeed, in appendix \ref{adiabapp} we construct an
explicit path $H(s)$ of this kind.

\section{Duality between spin models and string models} \label{dualsect}
In this section we explain the relationship between the spin Hamiltonians $H_0, H_1$, and
previously known models. Specifically, we show that $H_0, H_1$ are related via a duality map to
two previously studied lattice models -- the ``toric code'' model\cite{KitaevToric,Stringnet} and the
``doubled semion'' model.\cite{Stringnet}
The latter two models are sometimes called ``string models'' and are special cases of the general class of
``string-net'' models constructed in Ref. \onlinecite{Stringnet}. This duality provides another point of view on
the braiding statistics analysis in the previous section, and also suggests a natural classification scheme
for general 2D bosonic SPT phases with finite unitary symmetry groups.

\begin{figure}[t]
{\vskip -0.5cm \hspace*{0.0cm}\includegraphics[width=0.95\columnwidth]{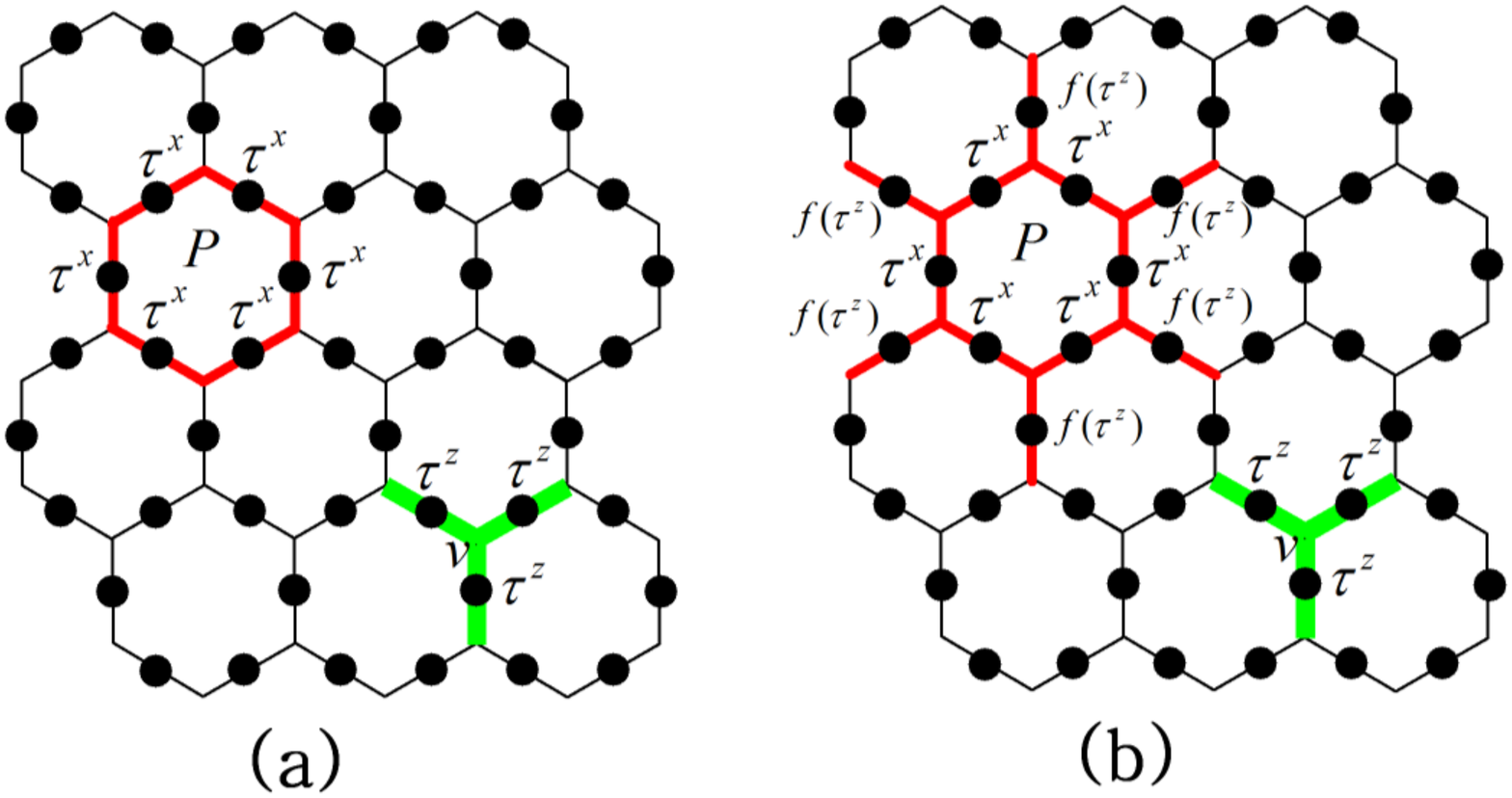}}\vskip -0.8cm
\caption{
The toric code and doubled semion models $H_{t.c}, H_{d.s}$ (\ref{Hdual}). In both systems, the Hilbert space is 
equivalent to a spin-$1/2$ model where the spins live on the links $l$ of the honeycomb lattice. 
(a) The toric code Hamiltonian $H_{t.c}$ is a sum of two terms. The first term, $Q_v$ (\ref{qvdef}) 
is a product of $\tau^z_l$ over the three links adjacent to the vertex $v$. The second term involves 
the interaction $\prod_{l \in p} \tau^x_l$ which acts on the six links adjacent to the plaquette $p$. 
(b) The doubled semion Hamiltonian $H_{d.s}$ includes the
same vertex term $Q_v$, but contains a more complicated plaquette term
$\prod_{l \in p} \tau^x_l \prod_{l \in \text{legs of } p} f(\tau^z_l)$ where $f(x) = i^{(1-x)/2}$.
} \label{fig:tc} 
\end{figure}

We begin by defining the duality map: we note that every spin configuration
$\{\sigma^z_p = \pm 1\}$ on the triangular lattice defines a corresponding domain wall configuration on the honeycomb
lattice. Formally, this correspondence is given by $\tau^z_l = \sigma^z_p \sigma^z_q$ where $l$ is the link separating
sites $p,q$ and $\tau^z_l = \mp 1$ corresponds to the presence or absence of a domain wall.
We will refer to these domain walls as ``strings.'' An important point is that the dual string degrees of freedom always form
closed loops -- that is, they satisfy the condition $Q_v = 1$ where (Fig. \ref{fig:tc})
\begin{equation}
Q_v = \prod_{l \in v} \tau^z_l
\label{qvdef}
\end{equation}

Using this correspondence, we can map our spin Hamiltonians $H_0, H_1$ (\ref{Hconv}-\ref{Hnew}) onto dual string Hamiltonians:
\begin{eqnarray}
H_0^{d} &=& - \sum_p   \left ( \prod_{l \in p} \tau^x_l \right)  \nonumber \\
H_1^{d} &=& \sum_p  \left( \prod_{l \in p} \tau^x_l \prod_{l \in \text{legs of } p} i^{\frac{1 - \tau^z_l}{2}} \right)
\label{H0d}
\end{eqnarray}
These Hamiltonians are defined on a Hilbert space consisting of \emph{closed} string states (i.e. states 
satisfying $Q_v = 1$ everywhere).

The dual Hamiltonians $H_0^{d}, H_1^{d}$ are closely related to two models studied in Ref. \onlinecite{Stringnet}:
the ``toric code model''\cite{KitaevToric} and the ``doubled semion'' model.
To understand the precise relationship, recall that the latter two models are defined on a
Hilbert space consisting of \emph{all} string states on the honeycomb lattice -- both open and closed. The two 
Hamiltonians are (Fig. \ref{fig:tc})
\begin{eqnarray}
H_{t.c} &=&  - \sum_v Q_v - \sum_p \left( \prod_{l \in p} \tau^x_l \right) P_p \nonumber \\
H_{d.s} &=&  - \sum_v Q_v +
\sum_p \left( \prod_{l \in p} \tau^x_l \prod_{l \in \text{legs of } p} i^{\frac{1 - \tau^z_l}{2}} \right) P_p
\nonumber\\\label{Hdual}
\end{eqnarray}
Here $P_p$ denotes the projector $P_p = \prod_{v \in p} (1+ Q_v)/2$. This operator defines a projection onto states
that satisfy the closed string constraint $Q_v = 1$ at all vertices of the plaquette $p$.

Comparing (\ref{Hdual}) and (\ref{H0d}), we see that $H_0^d, H_1^d$ can be obtained by restricting $H_{t.c}, H_{d.s}$
to the closed string ($Q_v = 1$) subspace. In other words, the spin models $H_0, H_1$
are dual to a restricted variant of the toric code and doubled semion models. 

In fact, this duality can be extended to one that maps the \emph{gauged} spin models
$\t{H}_0, \t{H}_1$ onto the \emph{unrestricted} toric code and doubled semion models (\ref{Hdual}).
The extended duality is defined by setting $\tau^z_l = \sigma^z_p \sigma^z_q \mu^z_{pq}, \tau^x_l = \mu^x_{pq}$
where $l$ is the link separating sites $p,q$. Substituting these expressions into
$H_{t.c}, H_{d.s}$ and making use of the gauge invariance constraint (\ref{gauss}) it is easy to check that the result is 
exactly $\t{H}_0, \t{H}_1$. We note that this duality maps local operators onto local (gauge invariant) operators 
and should therefore be thought of as an exact equivalence between two quantum systems. Thus the gauged spin models 
$\t{H}_0, \t{H}_1$ are physically identical to the toric code and doubled semion models. 

The above dualities are variants of the well-known correspondence between the 2D Ising model
and 2D $\mathbb{Z}_2$ gauge theory.\cite{Wegner,Kogut} To see this, note that the closed string models
$H_0^{d}, H_1^{d}$ are simply $\mathbb{Z}_2$ gauge theory Hamiltonians, phrased in the language of strings.
The Hamiltonian $H_0^{d}$ is the conventional\cite{Wegner,Kogut} $\mathbb{Z}_2$ gauge theory
Hamiltonian (in the zero coupling limit where there is no electric energy term $\sum_{l} \tau^z_l$),
while $H_1^{d}$ is another kind\cite{Wittencohomology} of $\mathbb{Z}_2$ gauge theory. From this point of view, the correspondence
between $H_0, H_1$ and $H_0^d, H_1^d$ is a duality between two types of 2D Ising paramagnets, and
two types of 2D $\mathbb{Z}_2$ gauge theory.

We can understand the duality between $\t{H}_0, \t{H}_1$
and $H_{t.c}, H_{d.s}$ in a similar way. We note that the first two models can be thought of as
two types of Ising paramagnets coupled to (conventional) $\mathbb{Z}_2$ gauge theory, while the latter two
models can be thought of as two types of $\mathbb{Z}_2$ gauge theory coupled to a (conventional) Ising paramagnet.
Hence the duality between $\t{H}_0, \t{H}_1$ and $H_{t.c}, H_{d.s}$ is a variant of
the well-known self-duality of 2D $\mathbb{Z}_2$ gauge theory coupled to Ising matter.\cite{Kogut}

We expect that these dualities can be generalized from $\mathbb{Z}_2$ to any finite unitary symmetry group $G$: each
SPT phase with symmetry group $G$ is dual to a corresponding gauge theory with gauge group $G$.
This correspondence immediately suggests a classification scheme for 2D bosonic SPT phases with finite unitary
symmetry groups: it is known that the different types of 2D gauge theories with group $G$ (or equivalently,
different string-net models corresponding to $G$) are in one-to-one correspondence with elements of $H^3(G, U(1))$.
(For a derivation of this result, see Ref. \onlinecite{Wittencohomology}, and also section 10.1.E.3 of 
Ref. \onlinecite{Kitaevhoneycomb}). Hence, the duality map
suggests that different SPT phases associated with symmetry group $G$ can also be classified by $H^3(G, U(1))$.
This classification scheme is identical to the proposal of Ref. \onlinecite{XieSPT4}. 

Another application of these dualities is that they give a simple method for constructing exactly soluble
models for bosonic SPT phases with finite unitary symmetry group $G$. The first step is to construct the different ``string-net''
models\cite{Stringnet} corresponding to the group $G$. These are models with string types given by the group elements $g \in G$, and branching
rules given by group multiplication: $\{g_1,g_2,g_3\}$ is an allowed branching if $g_1 g_2 g_3 = 1$. In general, there will be a
finite number of different models with these branching rules -- each one corresponding to a different solution of the self-consistency
equations of Ref. \onlinecite{Stringnet}.\footnote{For readers familiar with the string-net models in 
Ref. \onlinecite{Stringnet}, we note that one can equivalently construct string-net models by letting the string types be 
\emph{irreducible representations} of $G$, but these models are not as convenient for the duality construction when $G$ is 
non-abelian.}
We then take the dual of these models, and thereby construct exactly soluble models
for bosonic SPT phases. The models $H_0, H_1$ discussed here were constructed using this approach. In appendix \ref{toptermapp}
we show that an analogous duality in a space-time Lagrangian description can be used to construct topological
non-linear sigma models for SPT phases. 

\section{Protected edge modes and braiding statistics} \label{genargsect}
The most dramatic distinction between the two types of paramagnets
is that $H_1$ has protected gapless edge modes, while $H_0$ does not.
In other words, if we define $H_1$ in a geometry with a boundary, then the energy
spectrum always contains gapless excitations. These gapless excitations
are guaranteed to be present as long as the Ising symmetry is not broken
(explicitly or spontaneously). In this section, we give a general argument
proving this fact. Our argument reveals that these edge modes are closely connected
to the semionic braiding statistics of the $\pi$-flux excitations in the gauged spin model, $\t{H}_1$.
We note that the existence of protected edge modes was previously established in 
Ref. \onlinecite{XieSPT3} using a different approach.

The statement we prove is as follows. We consider a disk geometry with a Hamiltonian
of the form 
\begin{equation}
H = H_{\text{bulk}} + H_{\text{edge}} , \ \ \ H_{\text{bulk}} = - \sum_p B_p
\label{Htotal}
\end{equation}
where $B_p$ is defined as in (\ref{Hnew}) and the sum runs over all sites $p$ lying strictly in the interior
of the disk. We take the edge Hamiltonian $H_{\text{edge}}$ to be any Hamiltonian with local interactions
which acts on the spins on or near the boundary of the disk. In this setup, it is clear that the ground state $|\Psi\>$ 
of $H$ satisfies $B_p = 1$ when $p$ is far from the edge; in fact, in order to simplify the discussion, we will assume 
that $B_p = 1$ for all $p$ lying strictly in the interior of the disk. Given these assumptions, we will show that 
$|\Psi\>$ cannot be both Ising symmetric and short-range entangled. Here, a state is ``short-range entangled'' if it
can be transformed into a product state by a local unitary transformation -- a unitary operator generated from 
the time evolution of a local Hamiltonian over a finite time $t$.\cite{XieSPT1} 

To understand what this result means, recall that $|\Psi\>$ is always Ising symmetric and short-range entangled in the bulk 
(see appendix \ref{adiabapp}). Thus, the implication of the above theorem is that the edge either breaks the Ising symmetry or is not 
a short-range state. In the latter case, the edge is presumably gapless, so in this way we see that the edge is protected.

In section \ref{physintsect} we establish this result with an intuitive physical argument. In
section \ref{mathsect}, we give a rigorous mathematical proof. In section \ref{discsect}, we discuss 
generalizations to other systems.  

\begin{figure}[t]
{\includegraphics[width=0.9\columnwidth]{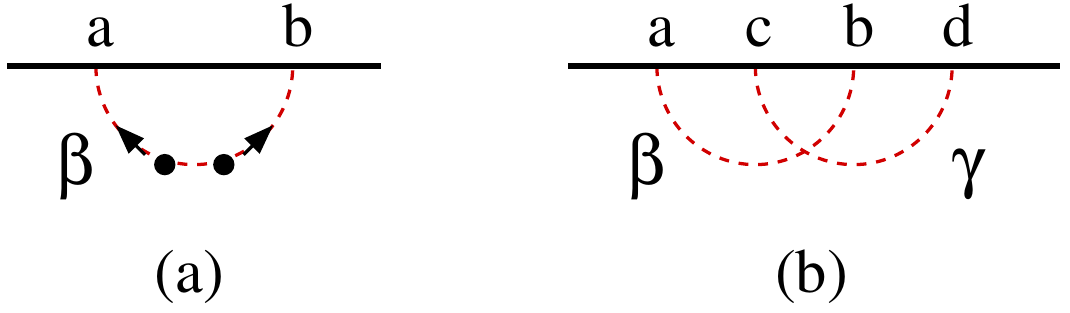}}
\caption{
(Color online) (a) We consider a process in which two $\pi$-fluxes are created in the bulk, 
moved to the boundary along a path $\beta$ and then annihilated near points $a,b$. 
(b) We prove that the edge is protected by considering two paths $\beta, \gamma$, and their corresponding 
flux creation/annihilation processes.
}\label{fig:braiding}
\end{figure}

\subsection{Physical argument} \label{physintsect}
The argument is a proof by contradiction: we assume that $|\Psi\>$ is both Ising                                                    
symmetric and short-range entangled and we show that these assumptions lead to a contradiction.
The first step is to consider a thought experiment in which we create a pair of $\pi$-fluxes in the bulk and
move them along some path $\beta$ to two points $a,b$ at the boundary (Fig. \ref{fig:braiding}a). 
This process can be implemented by applying an appropriate unitary operator to the state $|\Psi\>$.
We will denote this operator by $W_\beta$.
By construction $W_\beta |\Psi\>$ contains two fluxes located near points $a,b$ on the boundary.

We next assert that the $\pi$-fluxes at the boundary can be annihilated by local operators.
In other words, there exist local operators $U_a, U_b$ acting near $a,b$ such that
$U_a U_b W_\beta |\Psi\> = |\Psi\>$. To see this, note that the effect of bringing the $\pi$-flux excitations to the edge is to
create two Ising domain walls at points $a,b$. Given that $|\Psi\>$ is Ising symmetric
and short-range entangled, these domain walls are \emph{local} excitations -- that is, the two states, 
$|\Psi\>, W_\beta |\Psi\>$ have identical expectation values far from $a,b$. It then follows that these two states 
can be connected by local operators $U_a, U_b$ acting near these points. We emphasize that this conclusion depends crucially
on the Ising symmetry of $|\Psi\>$: if instead $|\Psi\>$ broke the Ising symmetry, the domain walls at $a,b$ would be nonlocal 
excitations, and there would be no way to annihilate them with local operators.

We now use the fact that $\pi$-fluxes can be annihilated at the boundary to derive a contradiction. Consider a three step process 
in which two $\pi$-fluxes are (1) created in the bulk, 
(2) moved to the boundary along the path $\beta$, and (3) annihilated. Let $\mathbb{W}_\beta$ be a unitary operator 
describing this process. (Formally, $\mathbb{W}_\beta$ is given by $\mathbb{W}_\beta \equiv U_a U_b W_\beta$). 
Consider a second path $\gamma$ with the geometry shown in Fig. \ref{fig:braiding}b, and define $\mathbb{W}_\gamma$ in the same 
way. By construction, we have $\mathbb{W}_\beta |\Psi\> = \mathbb{W}_\gamma |\Psi\> = |\Psi\>$. Hence,
\begin{equation}
\mathbb{W}_\beta \mathbb{W}_\gamma |\Psi\> = \mathbb{W}_\gamma \mathbb{W}_\beta |\Psi\> = |\Psi\>
\label{commrelphys}
\end{equation}
At the same time, it follows from general principles that $\mathbb{W}_\beta, \mathbb{W}_\gamma$ satisfy 
the commutation relation
\begin{equation}
\mathbb{W}_{\beta} \mathbb{W}_{\gamma} |\Psi\> = e^{2i\theta} \mathbb{W}_{\gamma} \mathbb{W}_{\beta} |\Psi\>
\label{statrelphys}
\end{equation}
where $\theta$ is the exchange statistics for the $\pi$-fluxes. (This result can be derived in the same way as Eq. (\ref{stat})).
To complete the argument, we note that the $\pi$-fluxes have semionic statistics so $e^{2i\theta} = -1$. 
Equations (\ref{commrelphys}),(\ref{statrelphys}) 
are therefore in contradiction, implying that our assumption must be false and the ground state $|\Psi\>$ cannot be both 
Ising symmetric and short-range entangled.

In this analysis, we have skated over an important subtlety. The issue is that we do not know whether $U_a, U_b$ 
are even or odd under the Ising symmetry. In other words, we do not know whether the flux annihilation process 
involves flipping an even or odd number of spins. To understand what this means, recall 
that there are actually two types of $\pi$-flux excitations which differ from one another by the addition of a 
spin-flip excitation $e$. Thus, the $U_a, U_b$ operators could describe the annihilation of either one of the two types of 
$\pi$-fluxes, depending on their parity. Since this parity is ambiguous, the existence of $U_a, U_b$ only
shows that \emph{at least one} of the two types of $\pi$-fluxes can be annihilated at the boundary. 

This subtlety becomes important in the last part of the argument where we derive a contradiction between equations
(\ref{commrelphys}),(\ref{statrelphys}). In particular, since we can only guarantee that one of the two types of $\pi$-fluxes can
be annihilated at the boundary, the proof is only valid if we show that these equations are inconsistent for \emph{both} types of fluxes.
Fortunately, this is not a problem: the two types of $\pi$-fluxes have exchange statistics $\theta = \pm \pi/2$,
so $e^{2i\theta} = -1$ in both cases. 

\subsection{Discussion and generalizations} \label{discsect}

The above argument does not use any properties of $H_1$ except the braiding statistics of the $\pi$-fluxes. Therefore, it
actually proves a more general statement: any $\mathbb{Z}_2$ SPT phase in which neither of the $\pi$-fluxes is a boson or a fermion 
is guaranteed to have a protected edge mode. Indeed, as long as $e^{2i\theta} \neq 1$ for both types of fluxes, 
the argument goes through unchanged. On the other hand, if either of the $\pi$-fluxes is a boson or a 
fermion -- as in a conventional paramagnet $H_0$ -- there is no contradiction between equations (\ref{commrelphys}), 
(\ref{statrelphys}) and the argument breaks down completely. From this point of view, the key reason that $H_1$ has a 
protected edge mode and $H_0$ doesn't, is the difference in their $\pi$-flux braiding statistics.

It is not hard to generalize the argument to arbitrary bosonic SPT phases with
unitary abelian symmetry groups $G$. For example, consider the case of $G = \mathbb{Z}_3$. Just as $\mathbb{Z}_2$ spin models support $\pi$-flux
excitations, models with $\mathbb{Z}_3$ symmetry support flux excitations with flux $2\pi/3$ and $4\pi/3$. These $2\pi/3$-fluxes and $4\pi/3$-fluxes
each come in three different types -- just like the two types of $\pi$-fluxes in the $\mathbb{Z}_2$ case. Using the same arguments as above, 
one can see that a $\mathbb{Z}_3$ SPT phase must have a protected edge unless there exists a set of two fluxes -- consisting of one $2\pi/3$-flux and 
one $4\pi/3$-flux -- such that (1) the fluxes in this set are bosons or fermions and (2) the fluxes in this set have trivial mutual statistics with respect to one another. Similarly to the $\mathbb{Z}_2$ case, this result can be derived by considering thought experiments where we
annihilate $2\pi/3$ and $4\pi/3$-fluxes at the boundary, and making use of the string commutation algebra (\ref{statrelphys}).
In fact, by using the statistical hopping algebra\cite{LevinWenHop} in place of (\ref{statrelphys}), 
we believe that this result can be strengthened even further: one can show the existence of a protected edge mode unless the above 
set of fluxes are all \emph{bosons}. We expect that similar generalizations exist for
the non-abelian case although we will not discuss them here.

\subsection{Mathematical argument} \label{mathsect}

Like the physical argument sketched above, the mathematical argument is a proof by contradiction. We assume that $|\Psi\>$ is both Ising symmetric and short range entangled 
(i.e. it can be turned into a product state by a local unitary transformation) and we show that these assumptions lead
to a contradiction.

\begin{figure}[t]
{\includegraphics[width=0.6\columnwidth]{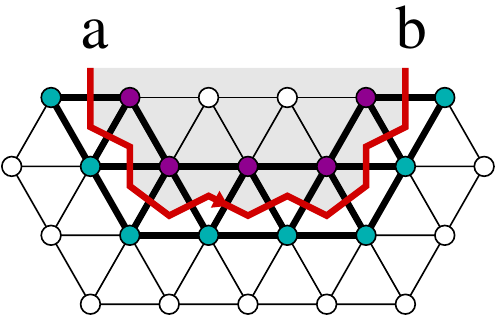}}
\caption{
(Color online) The operator $W_\beta$ (\ref{Wdef}) is defined for any path $\beta$ on the dual honeycomb lattice that joins points $a,b$
on the edge. In the interior of the path $\beta$ (shaded region), $W_{\beta}$ acts like the
symmetry transformation $S = \prod_p \sigma^x_p$. The operator also acts on all triangles $\<pqq'\>$ along the path $\beta$
(thickened lines). The action is different depending on whether $q,q'$ are to the left of $\beta$ (purple sites) or to the right
of $\beta$ (blue sites).
}\label{fig:wstr}
\end{figure}

To begin, let $\beta$ be a path on the dual (honeycomb) lattice that joins two points $a, b$
on the edge. We define an associated unitary operator $W_{\beta}$ by
\begin{equation}
W_{\beta} = \prod_{p, int} \sigma^x_p \cdot \prod_{\<pqq'\>, r} i^{\frac{1-\sigma^z_{q}\sigma^z_{q'}}{2}}
\cdot \prod_{\<pqq'\>,l} (-1)^{s_{pqq'}}
\label{Wdef}
\end{equation}
Here, the first product runs over all sites $p$ in the interior of the the path $\beta$, while the last two products
run over all triangles $\<pqq'\>$ along the path such that $q,q'$ are to the right of $\beta$
or to the left of $\beta$ respectively (Fig. \ref{fig:wstr}). The operator $s_{pqq'}$ is defined by
\begin{equation}
s_{pqq'} = \frac{1}{4}(1-\sigma^z_p \sigma^z_q)(1+\sigma^z_p \sigma^z_{q'})
\end{equation}

As an aside, we note that the operator $W_\beta$ is closely related to the string operator $V^1_\beta$ (\ref{string1}). Indeed the
two operators are identical except for the fact that $V^1_\beta$ is written in terms of the formalism of the gauged spin
model, while $W_\beta$ is written in terms of the original ``ungauged'' spin model. This similarity suggests a simple
physical interpretation for $W_\beta$: this operator describes a process in which two $\pi$-fluxes are created in 
the bulk and then moved along the path $\beta$ to points $a,b$ at the boundary. Much of what follows can be understood 
using this physical picture, as discussed in section \ref{physintsect}. 

Returning to the main argument, we note that the unitary operator $W_\beta$ has several important properties:
\begin{enumerate}
\item{ $W_{\beta}$ transforms local operators into local operators.
That is, $W_{\beta}^{-1} O W_{\beta}$ is local if and only if $O$ is local. \label{localprop} }
\item{ Let $O$ be a local operator which acts on spins within some convex region $R$
not containing either of the endpoints of $\beta$. Then $O$ has the same expectation value in the two states 
$|\Psi\>$ and $|\Psi'\> = W_\beta |\Psi\>$. \label{exprop} }
\end{enumerate}

Property \ref{localprop} follows from the fact that $W_\beta$ can be decomposed into a product
of two sets of commuting local unitary operators. As for property \ref{exprop}, there are three cases to consider:
the region of support $R$ may be contained entirely in the exterior of $\beta$, it may be contained entirely in the
interior, or it may overlap the path $\beta$ itself. In the first case, $W_{\beta}$ commutes with $O$, immediately
implying the desired equality $\<\Psi'| O |\Psi'\> = \<\Psi| O |\Psi\>$. In the second case,
$W_{\beta}^{-1} O W_{\beta} = S^{-1} O S$, since $W_\beta$ acts like $S$ in the interior of $\beta$. Then, since
$|\Psi\>$ is invariant under $S$ (\emph{by the Ising symmetry assumption}), we again have
$\<\Psi'| O |\Psi'\> = \<\Psi| O |\Psi\>$. The only case where the expectation value of $O$ could be different in
the two states is if $R$ overlaps the path $\beta$. However, one can check that $W_\beta |\Psi\> = W_{\beta'} |\Psi\>$ 
for any two paths $\beta, \beta'$ with the same endpoints.\footnote{This relation -- which is closely related to the
path independence of the string operator $V_\beta^1$ -- follows from the operator identity
$W_\beta = W_{\beta'} \prod_{p \in \beta' -\beta} B_p$, where the product runs over sites $p$ in between the
two paths $\beta, \beta'$.} This means that we can freely deform $\beta$ so that it 
avoids $R$. Therefore the expectation values must coincide in this case as well. 

We now use properties \ref{localprop}-\ref{exprop} to prove a key result: there exist local operators $U_a, U_b$ 
acting near $a,b$ (or more accurately, exponentially localized operators) such that
\begin{equation}
U_a U_b W_\beta |\Psi\> = |\Psi\>
\label{localann}
\end{equation}
The first step is to observe that $W_\beta |\Psi\>$ has short-range correlations (i.e., for any well
separated local operators $O_1, O_2$, we have $\<O_1 O_2\> = \<O_1\> \<O_2\>$
up to corrections which are exponentially small in the distance between $O_1, O_2$). To see this, note that
$|\Psi\>$ has short-range correlations since\cite{Hastingsadiab} it can be transformed into a product
state by a local unitary transformation (\emph{by the short-range entanglement assumption}). It then follows that 
$W_{\beta}|\Psi\>$ also has short range correlations since $W_{\beta}$ transforms local operators into local 
operators (property \ref{localprop}).

Next we recall that $|\Psi\>, W_\beta|\Psi\>$ share the same local expectation values away from the endpoints $a,b$ 
(property \ref{exprop}). Putting these facts together, we can
immediately deduce the existence of the desired $U_a, U_b$. To see this, consider the analogous question for the 
conventional paramagnet $|\Psi_0\> = |\sigma^x=1\>$: suppose that some short-range correlated state $|\Psi_0'\>$ 
has the same local expectation values as $|\Psi_0\>$ except near two points $a, b$. In this case, the state $|\Psi_0'\>$ must have
$\sigma^x = 1$ far from $a, b$, so it is clear that we can find local operators $U_a, U_b$ acting near $a,b$ such that
$U_a U_b |\Psi_0'\> = |\Psi_0\>$. Having established this property for $|\Psi_0\>$, it follows that the
same property must also hold for $|\Psi\>$ since $|\Psi\>,|\Psi_0\>$ are equivalent up to a local unitary transformation
(\emph{by the short-range entanglement assumption}).

A key question is to understand understand how $U_a, U_b$ transform under the Ising symmetry $S$. 
In appendix \ref{parityapp}, we show that $U_a, U_b$ can always be chosen so that they are either both even or both odd under $S$.
Furthermore, this even or odd parity must be the same for all pairs of endpoints $a,b$. In other
words, either all the $U_x$ operators are even under $S$, or all of them are odd under $S$.

We now use (\ref{localann}) to derive a contradiction. To this end, we consider a second path $\gamma$ that connects 
two other points $c, d$ on the edge. We choose $\beta, \gamma$ so that they intersect each other, and so that their endpoints are
well separated (see Fig. \ref{fig:braiding}b). As above, we have $U_c U_d W_{\gamma} |\Psi\> = |\Psi\>$ 
for some local operators $U_c, U_d$ acting near $c, d$. Now, define
\begin{align}
\mathbb{W}_{\beta} = U_a U_b W_{\beta}, \ \ \mathbb{W}_{\gamma} = U_c U_d W_{\gamma}
\end{align}
By construction, $\mathbb{W}_{\beta} |\Psi\> = |\Psi\>$ and
$\mathbb{W}_{\gamma} |\Psi\> = |\Psi\>$. Hence,
\begin{equation}
\mathbb{W}_{\beta} \mathbb{W}_{\gamma} |\Psi\> = \mathbb{W}_{\gamma} \mathbb{W}_{\beta} |\Psi\>
\label{commrel}
\end{equation}

At the same time, $\mathbb{W}_{\beta}, \mathbb{W}_{\gamma}$ \emph{anti-commute}, as we now show.
To see this, we first note that $W_\beta, W_\gamma$ anti-commute:
\begin{equation}
W_\beta W_\gamma = - W_\gamma W_\beta
\label{acomm}
\end{equation}
This relation can be checked using the explicit formula for $W_\beta$ (similarly to eq. (\ref{string1com})). Next, we 
recall that $W_{\beta}$ looks like $S$ in the interior of $\beta$ and the identity map
in the exterior of $\beta$ so that
\begin{equation}
W_{\beta} U_c U_d W_{\beta}^{-1} = (S U_c S^{-1}) U_d = \pm U_c U_d
\end{equation}
where the sign is determined by the parity of $U_c$ under $S$. Similarly, we have
\begin{equation}
W_{\gamma} U_a U_b W_{\gamma}^{-1} = U_a (S U_b S^{-1})  = \pm U_a U_b
\end{equation}
where the sign is determined by the parity of $U_b$ under $S$.
Importantly, these two signs are the same since the $U_x$ operators all share the same parity.
Hence, the two pairs $\{W_{\beta}, U_c U_d\}$ and $\{W_{\gamma}, U_a U_b\}$ either both commute
or both anti-commute. In either case, the anti-commutation relation (\ref{acomm}) implies
that $\mathbb{W}_{\beta}, \mathbb{W}_{\gamma}$ anti-commute:
\begin{equation}
\mathbb{W}_{\beta} \mathbb{W}_{\gamma} = - \mathbb{W}_{\gamma} \mathbb{W}_{\beta}
\label{anticommrel}
\end{equation}
Comparing (\ref{commrel}), (\ref{anticommrel}), we arrive at a contradiction.
Hence our assumption must be false and $|\Psi\>$ cannot be both Ising symmetric and short-range entangled.

\section{Microscopic edge analysis} \label{edgesect}
In this section, we investigate the protected edge modes of $H_1$ at a more
concrete level. We analyze a particular example of a gapless edge for $H_1$,
derive a field theoretic description of the low energy modes, and investigate the effect of perturbations.
As in section \ref{genargsect}, we consider a disk geometry, with a Hamiltonian of the form
$H = H_{\text{bulk}} + H_{\text{edge}}$. The bulk Hamiltonian $H_{\text{bulk}}$ is defined by $H_{\text{bulk}} = - \sum_p B_p$
where the sum runs over all sites that are strictly in the interior of the disk. The edge Hamiltonian
$H_{\text{edge}}$ can be any Ising symmetric Hamiltonian with local interactions which acts on the spins
on or near the boundary.

\subsection{Zero energy edge states}
We begin with the case where $H_{\text{edge}} = 0$ -- that is, the edge Hamiltonian vanishes. In this case,
we can compute the energy spectrum in the same way as we did for the periodic (torus) geometry.
First, we simultaneously diagonalize the $B_p$ operators for all sites $p$ that are strictly in the interior
of the disk. Next, we note that each of these simultaneous eigenstates is an energy eigenstate with energy
$E = - \sum_p b_p$ where $b_p = \pm 1$ is the eigenvalue under $B_p$. The final step is to determine the
degeneracy of these simultaneous eigenspaces. A natural guess, based on dimension counting,
is that each simultaneous eigenspace $\{b_p = \pm 1\}$ has a degeneracy of $2^N$, where $N$ is the number of spins along
the boundary of the disk. In particular, we expect that there are $2^N$ degenerate ground states.

\begin{figure}[t]
{\includegraphics[width=0.9\columnwidth]{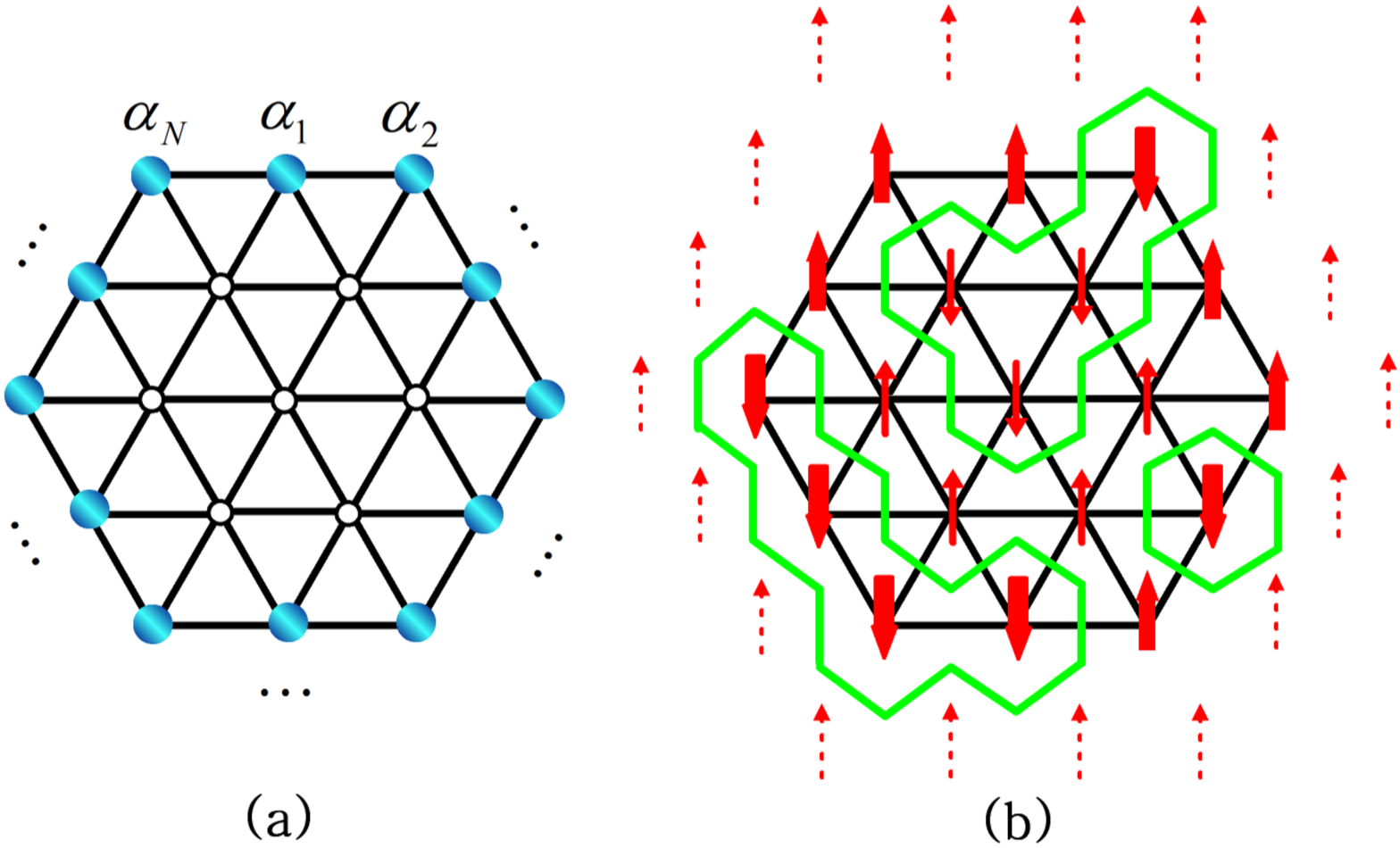}}
\caption{
A schematic picture of the zero energy edge states in the case $H_{\text{edge}}=0$.
(a) The $2^N$ zero energy edge states can be parameterized by
boundary spin configurations $\{\alpha_1,...,\alpha_N\}$, where $\alpha_n = \up,\down$.
(b) For each choice of $\{\alpha\}$, the corresponding wave function $\Psi_{\{\alpha\}}$
is defined by $\Psi_{\{\alpha\}}(\{\alpha_{int}\}) = (-1)^{N_{dw}}$ where $N_{dw}$ is the
total number of domain walls in the system. We use a convention where we close up all
the domain walls by assuming there is a ``ghost'' spin in the exterior of the disk, pointing
in the $\up$ direction.
}\label{fig:disc}
\end{figure}

We can verify this counting by constructing explicit wave functions for these degenerate ground states.
Specifically, we define a wave function $\Psi_{\{\alpha\}}$ for each boundary spin configuration $\{\alpha_1,...,\alpha_N\}$, 
where $\alpha_n = \up,\down$ (Fig. \ref{fig:disc}a). This wave function is a function
of the spins $\alpha_{int} = \uparrow,\downarrow$ lying strictly in the interior of the disk, and is given by
\begin{equation}
\Psi_{\{\alpha\}}(\{\alpha_{int}\}) = (-1)^{N_{dw}}
\label{edgestates}
\end{equation}
where $N_{dw}$ is the the total number of domain walls in the system.
Here, we define $N_{dw}$ using a particular convention where we close up all the
domain walls that end at the boundary by assuming that
there is a ``ghost'' spin in the exterior of the disk, pointing in the $\up$
direction (Fig. \ref{fig:disc}b). We will denote these states by $|\alpha_1,...,\alpha_N\>$.
As is apparent from this parameterization, we can think of these degenerate ground states as zero energy edge states.

It is useful to define operators $\{\wb{\sigma}^x_n, \wb{\sigma}^y_n, \wb{\sigma}^z_n\}$
that act on $|\alpha_1,...,\alpha_N\>$ just like the usual Pauli spin operators.
We note that the $\wb{\sigma}^i_n$ operators should not be confused with the physical boundary
spin operators $\sigma^i_n$ which act on the full Hilbert space of the spin system. In the $\sigma^z$ case,
the two types of operators are closely related -- for example, $\wb{\sigma}^z_n = P_0 \sigma^z_n P_0$
where $P_0$ is the projection operator onto the $2^N$ dimensional edge state subspace. However, this
simple relation does not hold for the $\sigma^x$ or $\sigma^y$ operators, or for more complicated products
of spin operators.

An important question is to understand how the symmetry $S$ acts on the edge states.
Using the definition (\ref{edgestates}), one finds that the Ising symmetry $S$ acts
as
\begin{equation}
S |\alpha_1,...,\alpha_N\> =  \pm \prod_{n=1}^N \sigma^x_{\alpha_n
\beta_n}|\beta_1,...,\beta_N\>
\end{equation}
where the sign depends on the configuration of $\alpha_n = \uparrow, \downarrow$ as follows: the sign is $-$
if the total number of domain walls between the $\alpha$'s is
divisible by $4$ and $+$ otherwise. In
other words, the action of the Ising symmetry on the above basis states
is described by the operator
\begin{equation}
S = -\prod_{n=1}^N \wb{\sigma}^x_n \cdot \exp\left(\frac{i\pi}{4}\sum_{n=1}^N
(1-\wb{\sigma}^z_n \wb{\sigma}^z_{n+1})\right)
\label{symmsigma}
\end{equation}

In order to gain some intuition about $S$, we note that the operators $\wb{\sigma}^x, \wb{\sigma}^y,
\wb{\sigma}^z$ transform under the symmetry according to
\begin{eqnarray}
S^{-1} \wb{\sigma}^x_n S &=& -\wb{\sigma}^z_{n-1} \wb{\sigma}^x_n \wb{\sigma}^z_{n+1} \nonumber \\
S^{-1} \wb{\sigma}^y_n S &=& \wb{\sigma}^z_{n-1} \wb{\sigma}^y_n \wb{\sigma}^z_{n+1} \nonumber \\
S^{-1} \wb{\sigma}^z_n S &=& -\wb{\sigma}^z_n
\end{eqnarray}

\subsection{An example of an edge Hamiltonian}
We now imagine adding a nonvanishing edge Hamiltonian $H_{\text{edge}}$. If $H_{\text{edge}}$ is small, then
we can analyze its effect using degenerate perturbation theory. The
first order splitting of the $2^N$ degenerate ground states can be obtained by diagonalizing
$P_0 H_{\text{edge}} P_0$ where $P_0$ is the projection onto the zero energy edge state subspace.
In general, $P_0 H_{\text{edge}} P_0$ can be expressed as a function of the
$\wb{\sigma}^i_n$ operators. We can therefore find the edge state spectrum by solving a $1D$ spin
chain with an unusual Ising symmetry (\ref{symmsigma}).\cite{XieSPT3}

\begin{figure}[t]
{\includegraphics[width=0.6\columnwidth]{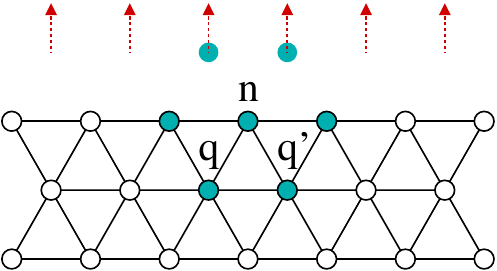}}
\caption{
The operator $B_n^\uparrow$ is defined just like $B_p$ (\ref{Hnew}), except with an additional
``ghost'' spin in the exterior of the disk pointing in the $\uparrow$ direction (dotted arrow).
More explicitly, $B_n^\uparrow$ acts on the three triangles $\<nqq'\>$ containing the boundary
spin $n$ with an action given by (\ref{Bndef}). The operator $B_n^\downarrow$ is defined similarly.
}\label{fig:Hedge}
\end{figure}

Here we will focus on a particular choice of $H_{\text{edge}}$ which can be solved exactly.
We emphasize that this choice is not unique, and that other edge Hamiltonians may give rise
to different edge spectra. Nevertheless, we believe that the particular $H_{\text{edge}}$ we consider
is a useful illustrative example. We will derive a low energy edge theory for this example,
and investigate the effect of perturbations.

Specifically, we consider an edge Hamiltonian
\begin{equation}
H_{\text{edge}} = - J \sum_{n=1}^N (B_n^\uparrow + B_n^\downarrow)
\label{edgeex}
\end{equation}
where $B_n^{\uparrow}$ is defined just like $B_p$, except with an
additional ``ghost'' spin in the exterior of the disk, pointing in
the $\uparrow$ direction. More explicitly,
\begin{equation}
B_n^\uparrow = -\sigma^x_n \cdot
i^{\frac{1-\sigma^z_{n-1}}{2}} \cdot i^{\frac{1-\sigma^z_{n+1}}{2}} \cdot \prod_{\<nqq'\>} i^{\frac{1-\sigma^z_{q}\sigma^z_{q'}}{2}}
\label{Bndef}
\end{equation}
where the product runs over the three triangles $\<nqq'\>$ containing the
boundary spin $n$, and where $n+1$, $n-1$ denote the two neighboring boundary
spins  (Fig. \ref{fig:Hedge}). The operator $B_n^\downarrow$ is defined the same way, except that
we take the ``ghost'' spin to point in the $\downarrow$ direction. That is,
\begin{equation}
B_n^\downarrow = -\sigma^x_n \cdot  i^{\frac{1+\sigma^z_{n-1}}{2}} \cdot i^{\frac{1+\sigma^z_{n+1}}{2}}
\cdot \prod_{\<nqq'\>} i^{\frac{1-\sigma^z_{q}\sigma^z_{q'}}{2}}
\end{equation}

The above edge Hamiltonian has several nice properties. First, the edge Hamiltonian
is Ising symmetric. Indeed, this follows from the fact that $S^{-1} B_n^\uparrow S = B_n^\downarrow$,
$S^{-1} B_n^\downarrow S = B_n^\uparrow$. Another property of $H_{\text{edge}}$ is that $[H_{\text{edge}}, H_{\text{bulk}}] = 0$.
This property follows from the fact that the $B_p$ operators commute with $B_n^\uparrow, B_n^\downarrow$ (which
in turn follows from the fact that the $B_p$ operators commute with each other). One consequence
of this commutation relation is that the low energy edge spectrum obtained by diagonalizing
$P_0 H_{\text{edge}} P_0$ is exact, rather than just being correct to first order in perturbation theory.

We now compute $P_0 H_{\text{edge}} P_0$. Using the definition of the basis states (\ref{edgestates}), it is easy to
check that
\begin{equation}
P_0 B_n^{\uparrow} P_0 = \wb{\sigma}^x_n
\end{equation}
from which it follows that
\begin{equation}
P_0 B_n^{\downarrow} P_0 =  S^{-1} \wb{\sigma}^x_n S = -\wb{\sigma}^z_{n-1} \wb{\sigma}^x_n \wb{\sigma}^z_{n+1}
\end{equation}
We conclude that:
\begin{equation}
P_0 H_{\text{edge}} P_0 = - J \sum_{n=1}^N (\wb{\sigma}^x_n - \wb{\sigma}^z_{n-1} \wb{\sigma}^x_n \wb{\sigma}^z_{n+1})
\label{edgeH1}
\end{equation}

A nice feature of this Hamiltonian is that it has a $U(1)$ symmetry:
it conserves $\sum_{n=1}^N \wb{\sigma}^z_n \wb{\sigma}^z_{n+1}$ --
the total number of domain walls between the boundary spins. In order
to make this $U(1)$ symmetry manifest and to simplify the analysis, it
is useful to rewrite the Hamiltonian in terms of the dual
domain wall variables. Naively, we can accomplish this by defining
\begin{equation}
\tau^z_n = \wb{\sigma}^z_{n} \wb{\sigma}^z_{n+1}
\end{equation}
and re-expressing everything in terms of the $\tau$'s.
However, the above duality transformation doesn't quite work for a system
with periodic boundary conditions, since the $\tau^z_n$ variables
obey the global constraint $\prod_{n=1}^N \tau^z_n = 1$, and therefore only
describe $N-1$ independent degrees of freedom. (Equivalently, there is no
way to express $\wb{\sigma}^z$ in terms of the $\tau^z$ variables).

In order to incorporate the missing degree of freedom
and make the dual description complete, we
introduce an additional $\mathbb{Z}_2$ gauge field
$\mu^z_{n-1,n}$ that lives on the links $\<(n-1) n\>$ connecting neighboring
boundary sites $(n-1), n$. We then
define the duality transformation between $\wb{\sigma}$ and $\tau, \mu$ by
the relation
\begin{equation}
\mu^x_{n-1,n} = \wb{\sigma}^z_n
\end{equation}
together with the gauge invariance constraint
\begin{equation}
\mu^x_{n-1,n} \mu^x_{n,n+1} \tau^z_n = 1
\label{giconst}
\end{equation}
It is easy to check that there
is a one-to-one correspondence between configurations of $\wb{\sigma}^z_n = \pm 1$
and configurations of $\mu^x_n = \pm 1, \tau^z_n = \pm 1$ obeying the
constraint (\ref{giconst}). Similarly, there is a one-to-one correspondence between
physical operators written in terms of the $\wb{\sigma}$'s and gauge invariant
combinations of $\mu, \tau$ (i.e. operators that commute with the left hand side of
(\ref{giconst})). In particular, the operators $\wb{\sigma}^x, \wb{\sigma}^y,
\wb{\sigma}^z$ are given by
\begin{eqnarray}
\wb{\sigma}^x_n &=& \tau^x_{n-1} \tau^x_n \mu^z_{n-1,n} \nonumber \\
\wb{\sigma}^y_n &=& -\tau^x_{n-1} \tau^x_n \mu^y_{n-1,n} \nonumber \\
\wb{\sigma}^z_n &=& \mu^x_{n-1,n}
\label{dualform}
\end{eqnarray}
while the symmetry transformation $S$ is given by
\begin{equation}
S = -\prod_{n=1}^N \mu^z_{n-1,n} \cdot \exp\left(\frac{i\pi}{4}\sum_{n=1}^N (1-\tau^z_n)\right)
\label{symmtau}
\end{equation}

Using (\ref{dualform}) to re-express the Hamiltonian (\ref{edgeH1}) in terms of the domain
wall variables $\tau$, we find
\begin{equation}
P_0 H_{\text{edge}} P_0 = -2J \sum_{n=1}^N (\tau^+_{n-1} \tau^-_n \mu^z_{n-1,n} + h.c.)
\label{edgetau}
\end{equation}
This Hamiltonian is the usual spin-$1/2$ XX chain, coupled to a $\mathbb{Z}_2$ gauge field
$\mu^z_{n-1,n}$. The only effect of the $\mathbb{Z}_2$ gauge field is to double the size of
the Hilbert space so that is includes sectors with both periodic and anti-periodic 
boundary conditions for the $\tau$ variables. The two types of boundary conditions 
correspond to the two possibilities $\prod_{n=1}^N \mu^z_{n-1,n} = \pm 1$.

\subsection{Edge theory}
Given previous work on the spin-$1/2$ XX chain, it is now straightforward to construct a field theory
description of the low energy edge modes. (We could also derive the \emph{exact} edge spectrum, but this is less
useful to us, as we ultimately want to analyze the effect of perturbations). To be specific, the low energy
excitations of the spin-$1/2$ XX chain (\ref{edgetau}) are known\cite{SubirBook} to be described by the non-chiral Luttinger liquid
\begin{eqnarray}
L &=& \frac{1}{4\pi} (\partial_x \theta \partial_t \phi + \partial_x
\phi \partial_t \theta) \nonumber \\
&-& \frac{v}{8\pi} \left( K (\partial_x \theta)^2 +
\frac{4}{K} (\partial_x \phi)^2 \right)\label{edgeth}
\end{eqnarray}
with Luttinger parameter $K = 1$, and velocity $v = 4Ja$ where $a$ is the lattice spacing.
Here, we are using a normalization convention in which expressions of the form
$e^{i k \theta + i l \phi}$ with integer $k,l$ correspond to local spin
operators (i.e. gauge invariant combinations of $\tau, \mu$).
For example,
\begin{eqnarray}
\tau^+_{n-1} \tau^+_n \mu^z_{n-1,n} &\sim& e^{i \theta} \nonumber \\
\mu^x_{n-1,n} &\sim& \cos(\phi) \nonumber \\
\frac{\tau^z_n}{2a} \sim \frac{1}{\pi} \partial_x \phi
\label{domwall}
\end{eqnarray}

In the above normalization convention, the boundary condition for $\theta$ is that
$\theta(L) \equiv \theta(0) \text{ (mod $2\pi$)}$. This condition automatically incorporates both the
periodic and anti-periodic sectors for $\tau$: the two sectors
correspond to the two cases $\theta(L) = \theta(0) + 4m\pi$ and $\theta(L) = \theta(0) + (4m+2)\pi$,
as one can see using the heuristic $\tau^+ \sim e^{i\theta/2}$.
The boundary condition for $\phi$ is also $\phi(L) \equiv \phi(0) \text{ (mod $2\pi$)}$.

The last component of the edge theory (\ref{edgeth}) is to understand how
the Ising symmetry $S$ acts in the new variables $\theta, \phi$. To this end,
we note that (\ref{domwall}) implies that
\begin{equation}
\exp\left(-\frac{\pi i}{4}\sum_{n=1}^N \tau^z_n\right) =
\exp\left(-\frac{i}{2} \int \partial_x \phi dx \right)
\end{equation}
Similarly, we have:
\begin{equation}
\prod_{n=1}^N \mu^z_{n-1,n} = \exp\left(\frac{i}{2} \int \partial_x \theta dx \right)
\end{equation}
This equality follows from the observation that the periodic/anti-periodic
sectors $\prod_{n=1}^N \mu^z_{n-1,n} = \pm 1$ correspond to the two boundary conditions
$\theta(L) - \theta(0) = 4m \pi, (4m+2)\pi$ respectively.

Combining these two results, we see that our expression (\ref{symmtau}) for $S$
becomes
\begin{equation}
S = \exp\left(\frac{i}{2} \int \partial_x \theta dx - \frac{i}{2} \int
\partial_x \phi dx \right)
\end{equation}
(up to a phase factor). Using the commutation relation $[\theta(x), \partial_x \phi(y)] =
2\pi i \delta(x-y)$, we deduce that
\begin{equation}
S^{-1} \theta S = \theta + \pi ;\quad S^{-1} \phi S = \phi + \pi
\label{symmedgeth}
\end{equation}
The transformation law (\ref{symmedgeth}) together with the
action (\ref{edgeth}) gives a complete description of the
low energy edge physics.

\subsection{Stability and instability of the edge modes}
In this section, we investigate the effect of perturbations on the gapless edge
(\ref{edgeth}). We find two results. Our first result is that the edge is \emph{unstable}: the edge modes can be
gapped out by arbitrarily small Ising symmetric perturbations. Our second result is that
the edge is \emph{protected}: we find that in all cases where perturbations gap out the edge, the Ising symmetry is
broken spontaneously. In other words, we do not find any perturbations which gap out the edge without breaking the Ising
symmetry, explicitly or spontaneously. This result is consistent with the general edge protection argument presented
in section \ref{genargsect}.

We focus on a particular class of perturbations of the form
\begin{equation}
U(l_1,l_2) = U(x) \cos(l_1\theta + l_2 \phi - \alpha(x))
\label{pert}
\end{equation}
where $l_1,l_2$ are integers. Using (\ref{symmedgeth}), we can see that the perturbation $U(l_1, l_2)$
is even or odd under the Ising symmetry depending on whether $l_1+l_2$ is even or odd, respectively.

The above perturbations are all ``local'' in the sense that they can be generated
by adding appropriate short range spin interactions
at the edge. For example, the case $U(0,1)$ can be generated by adding to the edge Hamiltonian
(\ref{edgeex}) a term of the form  $U \sigma^z_n$. Similarly, $U(1,0)$ can be generated by the
term $U (B_n^\uparrow - B_n^\downarrow)$.
Higher values of $l_1, l_2$ can be generated by more complicated spin interactions.

Of particular interest are perturbations of the form $U(l,0)$ and $U(0,l)$. We know from the standard
analysis of the sine-Gordon model that these terms can drive the edge to a gapped state by freezing
the value of $\theta$ or $\phi$. This gapping can occur even for infinitesimal $U$, if $U(l,0)$
or $U(0,l)$ are relevant in the renormalization group sense.

To determine whether any of these operators are relevant, we note that their scaling dimensions
are given by:
\begin{align}
h(l,0) = \frac{l^2}{K} ; \ \ h(0,l) = \frac{K l^2}{4}
\label{scal}
\end{align}
Clearly, the smaller the value of $l$, the more relevant the perturbation. On the
other hand, Ising symmetry (\ref{symmedgeth}) requires even $l$. Thus,
the most relevant Ising-symmetric operators are $U(2,0)$ and
$U(0,2)$. Setting $K=1$, we see that the term $U(2,0)$ has a scaling dimension greater than $2$
and is therefore irrelevant, but $U(0,2)$ is relevant. Hence, $U(0,2)$ describes an Ising-symmetric
instability of the edge.

Microscopically, the term $U(0,2)$ can be generated by adding a staggered spin interaction
\begin{equation}
U \sum_{n=1}^N (-1)^n (B_n^\uparrow + B_n^\downarrow)
\end{equation}
to $H_{\text{edge}}$ (\ref{edgeex}).
In this case, the resulting gapping of the edge modes can be analyzed exactly. The analysis is
similar to the derivation above: first, one maps the perturbed Hamiltonian onto an $XX$ chain with a staggered
coupling constant $J_n = J + (-1)^n U$. Then, one solves the resulting system using a Jordan-Wigner
transformation. One can check that the effect of the perturbation is to induce backscattering for
the non-interacting fermions, and hence open up a gap of order $U$.

Before proceeding further, we make two observations about this edge instability. The first
observation is that the instability described by $U(0,2)$ requires
the breaking of discrete translational symmetry. Indeed, as discussed in the previous paragraph,
$U(0,2)$ corresponds to backscattering between the left and right moving Jordan-Wigner fermions.
This backscattering process doesn't conserve the lattice momentum and therefore requires the breaking
of discrete translational symmetry and the doubling of the unit cell.

The second observation is that
the edge instability persists for any value of the Luttinger parameter, $K$. To see this, note that
(\ref{scal}) implies that $h(2,0) \cdot h(0,2) = 4$. There are three cases to consider: either
(a) $h(2,0) < 2 < h(0,2)$, (b) $h(0,2) < 2 < h(2,0)$, or (c) $h(2,0) = h(0,2) = 2$. In the first two cases, either
$U(2,0)$ or $U(0,2)$ is relevant, implying that the edge is unstable. In the third case, both operators
are marginal, but the edge is still unstable since small perturbations can affect $K$ and therefore make either $U(2,0)$
or $U(0,2)$ relevant. This analysis implies that the above edge theory has an Ising symmetric
instability for any value of $K$. (This instability is closely related to the fact that there is no
stable algebraic long range ordered phase in the 2D $\mathbb{Z}_4$ clock model\cite{WenBook}).

Although the perturbation $U(0,2)$ can open up a gap at the edge, it also spontaneously breaks the Ising
symmetry. To see this, note that $U(0,2)$ drives the edge into a state where $\phi$ is frozen
at some fixed value. In such a state, the operator $\cos(\phi-\alpha(x))$ acquires a nonvanishing
expectation value. But this operator is odd under $S$ (\ref{symmedgeth}) implying that the resulting state
spontaneously breaks the Ising symmetry. This result is consistent with the general argument in section \ref{genargsect}:
the edge modes can never be gapped out without breaking the Ising symmetry, either explicitly or spontaneously.

We have seen that the above edge is unstable in the sense that small perturbations can gap out the edge while
simultaneously breaking the Ising symmetry. We do not know whether a different choice of edge Hamiltonian $H_{\text{edge}}$
can give rise to a stable edge. Nevertheless, whether or not a stable $\mathbb{Z}_2$ edge is possible, we believe that the $\mathbb{Z}_n$
generalizations of the Ising paramagnet $H_1$ support stable gapless edge modes for $n>2$. Our expectation is based on the
following conjecture: we believe that the $\mathbb{Z}_n$ generalizations of $H_1$ support edge modes described by
(\ref{edgeth}) with a $\mathbb{Z}_n$ symmetry given by
\begin{equation}
S^{-1} \theta S = \theta + 2\pi/n ;\quad S^{-1} \phi S = \phi + 2\pi k/n
\label{symmedgethgen}
\end{equation}
with $k = 1, ..., n-1$. In this scenario, the most relevant $\mathbb{Z}_n$ symmetric perturbations are $U(n,0), U(0,n)$. Examining
(\ref{scal}), we can see that both of these operators are irrelevant over the finite range $8/n^2< K < n^2/2$.
Hence, if $K$ lies in this range, then the edge is stable to small perturbations.

\section{Conclusion}
In this paper we investigated a 2D bosonic SPT phase with a $\mathbb{Z}_2$
Ising-like symmetry. This SPT phase can be thought of as a new kind of Ising paramagnet. We showed
that this phase can be distinguished from a conventional paramagnet by coupling the system to
a $\mathbb{Z}_2$ gauge field and then analyzing the braiding statistics of the $\pi$-flux excitations. We found that
while the $\pi$-fluxes have bosonic or fermionic statistics in a conventional paramagnet, they have
semionic statistics in the new kind of paramagnet. This result immediately implies that the two types of
paramagnets belong to distinct phases. We also showed that these semionic braiding statistics directly imply the existence
of protected edge modes. To complete the picture, we analyzed a particular microscopic edge model for this phase, derived a
field theoretic description of the edge modes, and investigated their stability to perturbations.

While this paper has focused on a particular example, we believe that our basic approach can be applied more broadly.
The simplest extension would be to consider 2D bosonic SPT phases with arbitrary unitary symmetry groups $G$.
Following the same approach as above, we can couple each such SPT phase to a gauge field with
gauge group $G$. We can then analyze the quasiparticle excitations in this system
and find their braiding statistics. By analogy with the $\mathbb{Z}_2$ case, we expect that these braiding statistics can be used to uniquely
characterize each SPT phase and to derive the existence of protected edge modes. The same approach could potentially
be used for 2D fermionic SPT phases with unitary symmetries.

One can also imagine a generalization to higher dimensional bosonic/fermionic SPT phases with unitary symmetries. Again, we envision
coupling each phase to the appropriate gauge field and then analyzing the braiding statistics in the resulting system. In the 3D
case for example, we expect that gauged SPT phases will contain both particle-like and loop-like excitations. The analog of braiding
statistics is then the Berry phase associated with braiding a particle-like excitation around a loop-like excitation.
We find it plausible that these braiding statistics could be used to distinguish different SPT phases and to
derive the existence of protected boundary modes just as in 2D. Similarly, the duality between bosonic 2D SPT phases and
2D gauge theories (section \ref{dualsect}) may also extend to higher dimensions.

On the other hand, it is not clear how to apply these ideas to SPT phases with \emph{anti-unitary} symmetries
such as time reversal symmetry. The problem is that we do not know how to define a gauge field for an anti-unitary symmetry.
This question, as well as the potential generalizations discussed above, is an interesting direction for
future work.

\acknowledgments
ZCG thanks Xiao-Gang Wen, Alexei Kitaev and John Preskill for stimulating discussions. 
ZCG was supported in part by the NSF Grant No. NSFPHY05-51164. ML was supported in part by an 
Alfred P. Sloan Research Fellowship. 

\appendix

\section{Adiabatic equivalence of \texorpdfstring{$H_0, H_1$}{H0,H1} in the absence of the symmetry} \label{adiabapp}
While $H_0, H_1$ cannot be continuously connected when the Ising symmetry is preserved, these two models
can be connected when the symmetry is broken. Indeed, consider the one-parameter family of unitary transformations
\begin{equation}
U_\theta =
\prod_{\<pqr\>} e^{i \theta (3 \sigma^z_p \sigma^z_q \sigma^z_r - \sigma^z_p - \sigma^z_q - \sigma^z_r)}
\end{equation}
where the product runs over all triangles $\<pqr\>$.
Using these unitary transformations, we can define a one-parameter family of Hamiltonians
$H(\theta) =  U_\theta^{-1} H_0 U_\theta$. These Hamiltonians have local spin-spin interactions and have a finite energy
gap for any value of $\theta$. Moreover, $H(0) = H_0$, and one can check that $H(\pi/24) = H_1$, using the identity
\begin{equation}
U_{\pi/24}^{-1} \cdot \sigma^x_p \cdot U_{\pi/24} = B_p
\end{equation}
Hence, this construction gives an explicit path that connects $H_0, H_1$. We note that $H(\theta)$
breaks the Ising symmetry for intermediate values of $\theta$, as required by the argument of section \ref{pifluxsect}.

\section{Topological non-linear sigma models for the two paramagnets} \label{toptermapp}
In this section, we explore the duality between the spin models $H_0, H_1$ and the string models 
$H_{t.c.}, H_{d.s}$ in a space-time Lagrangian formulation. Using this space-time duality, 
we construct topological non-linear sigma models describing each of the two paramagnet phases, and thereby make a connection 
to the analysis of Ref. \onlinecite{XieSPT4}.

We begin with the Lagrangian description of the toric code and doubled semion models, $H_{t.c}, H_{d.s}$.
These models -- like all string-net models\cite{Stringnet} -- have a Euclidean space-time description in
terms of Turaev-Viro\cite{TV} invariants. In general, these invariants define a space-time partition function
for any 3D manifold $M$ and any triangulation of $M$ into tetrahedra. For the above two models
the Turaev-Viro invariants are of the form
\begin{equation}
Z=\frac{1}{2^{N_v}}\sum_{ijk\ldots} \prod_{\rm{link}}d_i
\prod_{\rm{tetrahedron}} G_{kln}^{ijm}, \label{partition}
\end{equation}
where the degrees of freedom $i,j,k...$ live on the links of the tetrahedra and run over the finite set $\{0,1\}$.
The variable $N_v$ denotes the number of vertices in the triangulation.

\begin{figure}[t]
{\vskip -0.5cm\includegraphics[width=0.9\columnwidth]{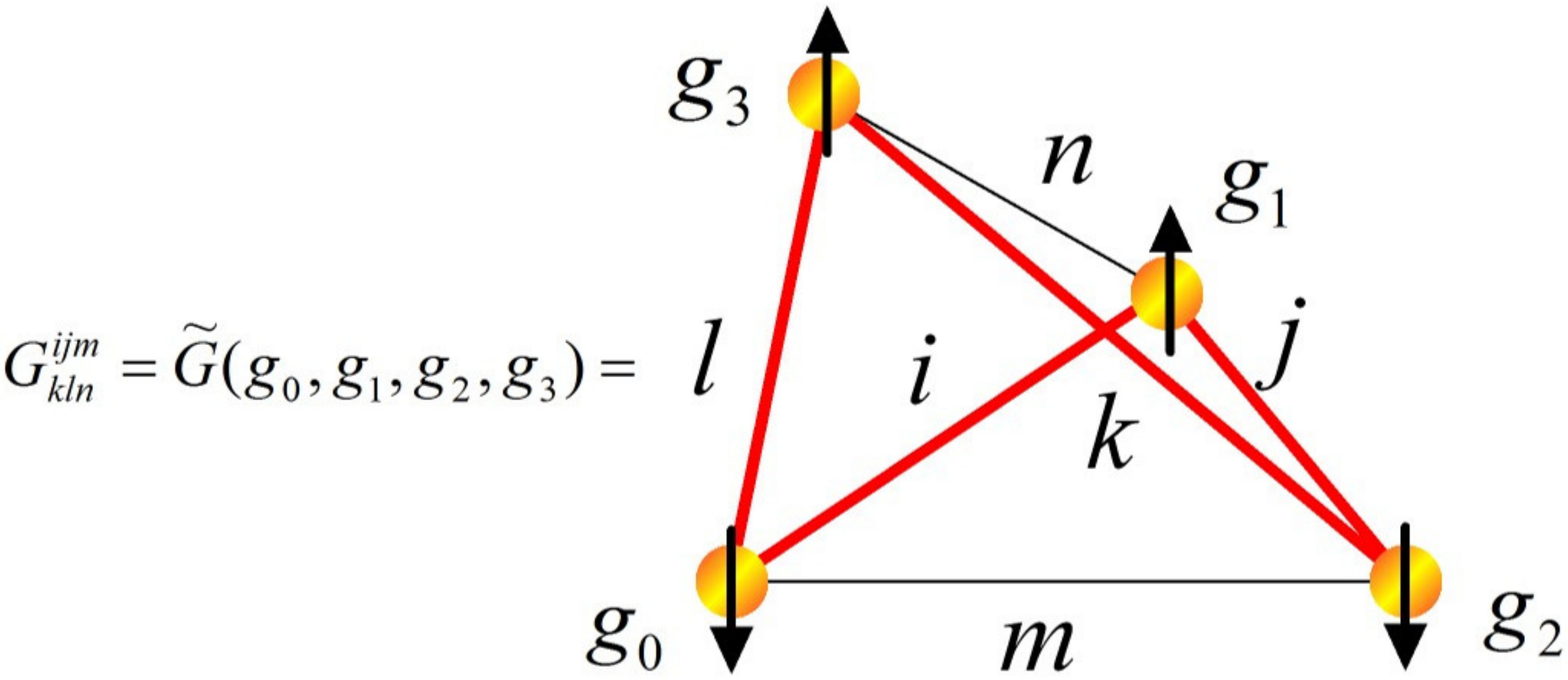}}\vskip -1.0cm 
\caption{
(Color online) In the Turaev-Viro model (\ref{partition}), the degrees of freedom $i,j,k = 0,1$ live on the links of the 
triangulation and $G_{kln}^{ijm}$ gives a weight to each tetrahedron in the space-time triangulation.
In the dual spin model, the degrees of freedom are Ising spins living on the vertices of the triangulation while
the dual weight $\t G$ is defined by mapping domain walls between spins onto the link variables $i=0,1$.
For example, the above configuration corresponds to $\t G(\down,\up,\down,\up)=G^{110}_{110}$. The thick
red lines denote links with $i=1$. }\label{fig:dual}
\end{figure}

To specify the two partition functions, we need to define $d_i, G_{kln}^{ijm}$ -- the weights associated
with the links and tetrahedra in the triangulation. For toric code model, we have $d_0=d_1=1$ and\cite{Stringnet}
\begin{eqnarray}
G_{000}
^{000}&=&1, \nonumber\\
G^{000}
_{111} &=& G^{110}
_{001} = G^{011}
_{100} = G^{101}
_{010}=1, \nonumber\\
G^{110}
_{110} &=& G^{101}
_{101} = G^{011}
_{011}=1, \nonumber\\
\rm{others}&=&0 \label{Gtc}
\end{eqnarray}
For double semion model, we have $d_0=1, d_1=-1$ and
\begin{eqnarray}
G_{000}
^{000}&=&1, \nonumber\\
G^{000}
_{111} &=& G^{110}
_{001} = G^{011}
_{100} = G^{101}
_{010}=-i, \nonumber\\
G^{110}
_{110} &=& G^{101}
_{101} = G^{011}
_{011}=-1, \nonumber\\
\rm{others}&=&0 \label{Gds}
\end{eqnarray}
Our convention for ordering the indices of $G$ is that the $3$ upper indices $i,j,m$ live on one of the faces of the
tetrahedron, while the corresponding lower indices $k,l,n$ live on the opposite links (Fig. \ref{fig:dual}).
There is no further ambiguity in the index ordering since the $G$-symbols for these models have full tetrahedral symmetry.

An important property of these partition functions is that they are independent of the choice of triangulation,
and depend only on the topology of the space-time manifold $M$. This triangulation independence should not be 
taken for granted: it only comes about because $G^{ijm}_{kln}, d_i, D$ satisfy highly nontrivial algebraic relations.\cite{TV}

We are now ready to discuss the space-time description of the duality between the spin models $H_0, H_1$ and 
the string models $H_{t.c}, H_{d.s}$. By analogy with the Hamiltonian description (section \ref{dualsect}),
we place the dual Ising spins on the vertices of the tetrahedra and then define the duality by mapping each Ising spin configuration 
onto its corresponding domain wall configuration (Fig. \ref{fig:dual}). To be precise, given any Ising spin configuration, we define 
a corresponding configuration of $i=0,1$ by placing $i=1$ on the links where the adjoining spins are anti-parallel 
(i.e. where there is a domain wall) and $i=0$ on the links where the spins are parallel
(i.e. where there is no domain wall). Importantly, we can see from (\ref{Gtc},\ref{Gds}) that 
$G$ vanishes for configurations of $i,j,k...$ which do not correspond to valid domain wall configurations, 
so this correspondence is (locally) one-to-one. In this way, we can map the two Turaev-Viro models (\ref{partition})
onto two spin partition functions
\begin{equation}
\t Z=\frac{1}{2^{N_v}}\sum_{g_0g_1g_2g_3\ldots} \prod_{\rm{link}}\t d(g_0,g_1)
\prod_{\rm{tetrahedron}} \t G(g_0,g_1,g_2,g_3),
\label{dualpartition}
\end{equation}
where $g_i=\up,\down$ runs over the two Ising spin states. The dual $G$-symbols $\t G(g_0,g_1,g_2,g_3)$ are defined by
\begin{eqnarray}
\t G(\up,\up,\up,\up)&=&\t G(\down,\down,\down,\down)=G^{000}_{000},\nonumber\\
\t G(\up,\up,\up,\down)&=&\t G(\down,\down,\down,\up)=G^{000}_{111},\nonumber\\
\t G(\up,\down,\up,\up)&=&\t G(\down,\up,\down,\down)=G^{110}_{001},\nonumber\\
\t G(\up,\up,\down,\up)&=&\t G(\down,\down,\up,\down)=G^{011}_{100},\nonumber\\
\t G(\down,\up,\up,\up)&=&\t G(\up,\down,\down,\down)=G^{101}_{010},\nonumber\\
\t G(\up,\down,\up,\down)&=&\t G(\down,\up,\down,\up)=G^{110}_{110},\nonumber\\
\t G(\up,\down,\down,\up)&=&\t G(\down,\up,\up,\down)=G^{101}_{101},\nonumber\\
\t G(\up,\up,\down,\down)&=&\t G(\up,\up,\down,\down)=G^{011}_{011}.
\end{eqnarray}
while $\t d(g_0,g_1)$ is given by
\begin{eqnarray}
\t d(\up,\up)&=&\t d(\down,\down)=d_0 \nonumber\\
\t d(\up,\down)&=&\t d(\down,\up)=d_1,
\end{eqnarray}

By construction, $\t G$ and $\t d$ are invariant under Ising symmetry:
\begin{eqnarray}
 \t G(g g_0,g g_1,g g_2,g g_3)&=& \t G(g_0,g_1,g_2,g_3), \nonumber\\ \t d(g g_0,g g_1)&=& \t d(g_0,g_1)
\end{eqnarray}
where $g \cdot \up = \down, g \cdot \down = \up$.
Thus, the dual partition functions (\ref{dualpartition}) both describe Ising
symmetric phases. We expect that these two phases correspond to the two types of paramagnets, $H_0, H_1$.

In addition, we note that the dual partition functions (\ref{dualpartition}) satisfy the property that they 
are independent of the choice of triangulation: this result follows from the corresponding property of the 
Turaev-Viro partition function. This property suggests that the two actions described by (\ref{dualpartition}) can be regarded as
$\mathbb{Z}_2$ topological non-linear sigma models similar to those constructed in Ref. \onlinecite{XieSPT4}.

We expect that this construction of topological non-linear sigma models can be generalized to arbitrary bosonic
2D SPT phases with finite unitary symmetry group $G$. The first step is to find all the 
Turaev-Viro models associated with the group $G$. In these Turaev-Viro models -- which are equivalent
to the topological gauge theories discussed in Ref. \onlinecite{Wittencohomology} -- 
the link labels $i$ run over the different \emph{group elements} of $G$. (One can 
also construct Turaev-Viro models by placing \emph{irreducible representations} of $G$ on the links, but this
approach is less convenient here). One can then construct dual models by placing group elements 
(``spins'') on the vertices of the tetrahedra, and mapping the domain walls between these generalized spins 
onto the group elements living on the links. The result will be a set of topological non-linear sigma models with
symmetry group $G$. 

\section{Parity of \texorpdfstring{$U_a, U_b$}{Ua,Ub} under \texorpdfstring{$S$}S}\label{parityapp}
In this section, we show that the operators $U_a, U_b$ defined by
$U_a U_b W_\beta |\Psi\> = |\Psi\>$ can always be chosen so that $U_a, U_b$
are either both even or both odd under $S$. Furthermore, we show that
this even or odd parity must be the same for all pairs of endpoints $a,b$.

We begin with the first claim -- showing that $U_a, U_b$ can always be chosen so that
they are either both even or both odd under $S$. To derive this fact, we define even and odd combinations
$U_{a\pm} = \frac{1}{2}(U_a \pm S^{-1} U_a S)$, and similarly for $U_b$.
We then have
\begin{equation}
|\Psi\> = (U_{a+}+U_{a-}) (U_{b+} + U_{b-}) |\Psi'\>
\end{equation}
where $|\Psi'\> = W_\beta |\Psi\>$.
We note that $|\Psi\>, |\Psi'\>$ have the same parity under $S$ since $W_{\beta}$ is even under $S$.
It then follows from symmetry that
\begin{equation}
U_{a+} U_{b-}|\Psi'\> + U_{a-} U_{b+} |\Psi'\> = 0
\end{equation}
On the other hand, it is easy to see that the two states
$U_{a+}U_{b-} |\Psi'\>, U_{a-}U_{b+} |\Psi'\>$ must be orthogonal to one another:
\begin{eqnarray}
\<\Psi'|U_{b-}^\dagger U_{a+}^\dagger U_{a-} U_{b+} |\Psi'\> &=&
\<\Psi'|(U_{a+}^\dagger U_{a-}) (U_{b-}^\dagger U_{b+}) |\Psi'\> \nonumber \\
&=&
\<\Psi'|U_{a+}^\dagger U_{a-} |\Psi'\> \nonumber \\
&\cdot& \<\Psi'|U_{b-}^\dagger U_{b+} |\Psi'\> \nonumber \\
&=& 0
\end{eqnarray}
where the second equality follows from the fact that $|\Psi'\>$ has short range correlations,
and the last equality follows from the fact that $|\Psi'\>$ has a definite parity under $S$.
Given that the two states are orthogonal and sum to zero, both states must vanish:
\begin{equation}
U_{a+} U_{b-}|\Psi'\> = U_{a-}U_{b+} |\Psi'\> = 0
\end{equation}
Next, we use the fact that $|\Psi'\>$ has short range correlations to deduce that
\begin{eqnarray}
0 &=& \<\Psi'| U_{b-}^\dagger U_{a+}^\dagger U_{a+} U_{b-} |\Psi'\> \nonumber \\
&=& \<\Psi'|  (U_{a+}^\dagger U_{a+}) (U_{b-}^\dagger U_{b-}) |\Psi'\> \nonumber \\
&=& \<\Psi'|  (U_{a+}^\dagger U_{a+}) |\Psi\>\<\Psi|(U_{b-}^\dagger U_{b-}) |\Psi'\>
\end{eqnarray}
Hence, either $U_{a+}|\Psi'\> = 0$ or $U_{b-} |\Psi'\> = 0$. In the same way,
we can show that either $U_{a-}|\Psi'\> = 0$ or $U_{b+} |\Psi'\> = 0$. There are
only two consistent possibilities: either $U_{a-}|\Psi'\> = U_{b-}|\Psi'\> = 0$
or $U_{a+}|\Psi'\> = U_{b+}|\Psi'\> = 0$. In the first case, we can replace $U_a \rightarrow U_{a+},
U_{b} \rightarrow U_{b+}$ so that both operators are even under $S$. Similarly, in the second case, we can
replace $U_a \rightarrow U_{a-}, U_b \rightarrow U_{b-}$ so that both operators are odd under $S$.
This establishes the first claim.

We now prove the second claim -- i.e. that this even or odd parity is the same for all endpoints $a,b$.
To see this, let $\beta$ be a path joining $a,b$ and $\beta'$ be a path joining $a$
with some other point $c$. Then, we have $U_a U_b W_{\beta}|\Psi\> = |\Psi\>$ and
$U'_a U'_c W_{\beta'}|\Psi\> = |\Psi\>$ for some
operators $U_a, U_b, U'_a, U'_c$. Now, by the result above, we know that $U_a, U_b$ have the same parity
and $U'_a, U'_c$ have the same parity. Also, it is not hard to see that $U'_a, U_a$ have the same parity:
in fact, we can always choose $U'_a = U_a$ up to a phase factor, since $W_{\beta'}|\Psi\>, W_{\beta}|\Psi\>$ have
the same local expectation values near $a$. It then follows that $U_b, U'_c$ also
have the same parity under $S$.

We now repeat this argument for a path $\beta''$ connecting $c$ with
some other point $d$, letting $ U''_c U''_d W_{\beta''} |\Psi\> = |\Psi\>$. By the same reasoning,
we find that $U_a, U_b, U''_c, U''_d$ all have the same parity under $S$ for arbitrary points $a,b,c,d$.
This establishes the claim.

\bibliography{duality}

\begin{thebibliography}{10}%
\makeatletter
\providecommand \@ifxundefined [1]{%
 \ifx #1\undefined \expandafter \@firstoftwo
 \else \expandafter \@secondoftwo
\fi
}%
\providecommand \@ifnum [1]{%
 \ifnum #1\expandafter \@firstoftwo
 \else \expandafter \@secondoftwo
\fi
}%
\providecommand \enquote [1]{``#1''}%
\providecommand \bibnamefont  [1]{#1}%
\providecommand \bibfnamefont [1]{#1}%
\providecommand \citenamefont [1]{#1}%
\providecommand\href[0]{\@sanitize\@href}%
\providecommand\@href[1]{\endgroup\@@startlink{#1}\endgroup\@@href}%
\providecommand\@@href[1]{#1\@@endlink}%
\providecommand \@sanitize [0]{\begingroup\catcode`\&12\catcode`\#12\relax}%
\@ifxundefined \pdfoutput {\@firstoftwo}{%
 \@ifnum{\z@=\pdfoutput}{\@firstoftwo}{\@secondoftwo}%
}{%
 \providecommand\@@startlink[1]{\leavevmode\special{html:<a href="#1">}}%
 \providecommand\@@endlink[0]{\special{html:</a>}}%
}{%
 \providecommand\@@startlink[1]{%
  \leavevmode
  \pdfstartlink
   attr{/Border[0 0 1 ]/H/I/C[0 1 1]}%
   user{/Subtype/Link/A<</Type/Action/S/URI/URI(#1)>>}%
  \relax
 }%
 \providecommand\@@endlink[0]{\pdfendlink}%
}%
\providecommand \url  [0]{\begingroup\@sanitize \@url }%
\providecommand \@url [1]{\endgroup\@href {#1}{\urlprefix}}%
\providecommand \urlprefix [0]{URL }%
\providecommand \Eprint[0]{\href }%
\@ifxundefined \urlstyle {%
  \providecommand \doi [1]{doi:\discretionary{}{}{}#1}%
}{%
  \providecommand \doi [0]{doi:\discretionary{}{}{}\begingroup
  \urlstyle{rm}\Url }%
}%
\providecommand \doibase [0]{http://dx.doi.org/}%
\providecommand \Doi[1]{\href{\doibase#1}}%
\providecommand \bibAnnote [3]{%
  \BibitemShut{#1}%
  \begin{quotation}\noindent
    \textsc{Key:}\ #2\\\textsc{Annotation:}\ #3%
  \end{quotation}%
}%
\providecommand \bibAnnoteFile [2]{%
  \IfFileExists{#2}{\bibAnnote {#1} {#2} {\input{#2}}}{}%
}%
\providecommand \typeout [0]{\immediate \write \m@ne }%
\providecommand \selectlanguage [0]{\@gobble}%
\providecommand \bibinfo [0]{\@secondoftwo}%
\providecommand \bibfield [0]{\@secondoftwo}%
\providecommand \translation [1]{[#1]}%
\providecommand \BibitemOpen[0]{}%
\providecommand \bibitemStop [0]{}%
\providecommand \bibitemNoStop [0]{.\EOS\space}%
\providecommand \EOS [0]{\spacefactor3000\relax}%
\providecommand \BibitemShut [1]{\csname bibitem#1\endcsname}%
\bibitem{KaneMele}%
  \BibitemOpen
  \bibfield{author}{%
  \bibinfo {author} {\bibfnamefont{C.~L.}\ \bibnamefont{Kane}}\ and\ \bibinfo
  {author} {\bibfnamefont{E.~J.}\ \bibnamefont{Mele}},\ }%
  \bibfield{journal}{%
  \Doi{10.1103/PhysRevLett.95.226801}{\bibinfo {journal} {Phys. Rev. Lett.}}\
  }%
  \textbf{\bibinfo {volume} {95}},\ \bibinfo {pages} {226801} (\bibinfo {year}
  {2005})%
  \bibAnnoteFile{NoStop}{KaneMele}%
\bibitem{KaneMele2}%
  \BibitemOpen
  \bibfield{author}{%
  \bibinfo {author} {\bibfnamefont{C.~L.}\ \bibnamefont{Kane}}\ and\ \bibinfo
  {author} {\bibfnamefont{E.~J.}\ \bibnamefont{Mele}},\ }%
  \bibfield{journal}{%
  \Doi{10.1103/PhysRevLett.95.226801}{\bibinfo {journal} {Phys. Rev. Lett.}}\
  }%
  \textbf{\bibinfo {volume} {95}},\ \bibinfo {pages} {146802} (\bibinfo {year}
  {2005})%
  \bibAnnoteFile{NoStop}{KaneMele2}%
\bibitem{Roy}%
  \BibitemOpen
  \bibfield{author}{%
  \bibinfo {author} {\bibfnamefont{R.}~\bibnamefont{Roy}},\ }%
  \bibfield{journal}{%
  \Doi{10.1103/PhysRevB.79.195322}{\bibinfo {journal} {Phys. Rev. B}}\ }%
  \textbf{\bibinfo {volume} {79}},\ \bibinfo {pages} {195322} (\bibinfo {year}
  {2009})%
  \bibAnnoteFile{NoStop}{Roy}%
\bibitem{FuKaneMele}%
  \BibitemOpen
  \bibfield{author}{%
  \bibinfo {author} {\bibfnamefont{L.}~\bibnamefont{Fu}}, \bibinfo {author}
  {\bibfnamefont{C.~L.}\ \bibnamefont{Kane}},\ and\ \bibinfo {author}
  {\bibfnamefont{E.~J.}\ \bibnamefont{Mele}},\ }%
  \bibfield{journal}{%
  \Doi{10.1103/PhysRevLett.98.106803}{\bibinfo {journal} {Phys. Rev. Lett.}}\
  }%
  \textbf{\bibinfo {volume} {98}},\ \bibinfo {pages} {106803} (\bibinfo {year}
  {2007})%
  \bibAnnoteFile{NoStop}{FuKaneMele}%
\bibitem{MooreBalents}%
  \BibitemOpen
  \bibfield{author}{%
  \bibinfo {author} {\bibfnamefont{J.~E.}\ \bibnamefont{Moore}}\ and\ \bibinfo
  {author} {\bibfnamefont{L.}~\bibnamefont{Balents}},\ }%
  \bibfield{journal}{%
  \Doi{10.1103/PhysRevB.75.121306}{\bibinfo {journal} {Phys. Rev. B}}\ }%
  \textbf{\bibinfo {volume} {75}},\ \bibinfo {pages} {121306} (\bibinfo {year}
  {2007})%
  \bibAnnoteFile{NoStop}{MooreBalents}%
\bibitem{HasanKaneRMP}%
  \BibitemOpen
  \bibfield{author}{%
  \bibinfo {author} {\bibfnamefont{M.~Z.}\ \bibnamefont{Hasan}}\ and\ \bibinfo
  {author} {\bibfnamefont{C.~L.}\ \bibnamefont{Kane}},\ }%
  \bibfield{journal}{%
  \bibinfo {journal} {Rev. Mod. Phys.}\ }%
  \textbf{\bibinfo {volume} {82}},\ \bibinfo {pages} {3045} (\bibinfo {year}
  {2010})%
  \bibAnnoteFile{NoStop}{HasanKaneRMP}%
\bibitem{GuSPT}%
  \BibitemOpen
  \bibfield{author}{%
  \bibinfo {author} {\bibfnamefont{Z.-C.}\ \bibnamefont{Gu}}\ and\ \bibinfo
  {author} {\bibfnamefont{X.-G.}\ \bibnamefont{Wen}},\ }%
  \bibfield{journal}{%
  \Doi{10.1103/PhysRevB.80.155131}{\bibinfo {journal} {Phys. Rev. B}}\ }%
  \textbf{\bibinfo {volume} {80}},\ \bibinfo {pages} {155131} (\bibinfo {year}
  {2009})%
  \bibAnnoteFile{NoStop}{GuSPT}%
\bibitem{XieSPT1}%
  \BibitemOpen
  \bibfield{author}{%
  \bibinfo {author} {\bibfnamefont{X.}~\bibnamefont{Chen}}, \bibinfo {author}
  {\bibfnamefont{Z.-C.}\ \bibnamefont{Gu}},\ and\ \bibinfo {author}
  {\bibfnamefont{X.-G.}\ \bibnamefont{Wen}},\ }%
  \bibfield{journal}{%
  \Doi{10.1103/PhysRevB.83.035107}{\bibinfo {journal} {Phys. Rev. B}}\ }%
  \textbf{\bibinfo {volume} {83}},\ \bibinfo {pages} {035107} (\bibinfo {year}
  {2011})%
  \bibAnnoteFile{NoStop}{XieSPT1}%
\bibitem{XieSPT2}%
  \BibitemOpen
  \bibfield{author}{%
  \bibinfo {author} {\bibfnamefont{X.}~\bibnamefont{Chen}}, \bibinfo {author}
  {\bibfnamefont{Z.-C.}\ \bibnamefont{Gu}},\ and\ \bibinfo {author}
  {\bibfnamefont{X.-G.}\ \bibnamefont{Wen}},\ }%
  \bibfield{journal}{%
  \Doi{10.1103/PhysRevB.84.235128}{\bibinfo {journal} {Phys. Rev. B}}\ }%
  \textbf{\bibinfo {volume} {84}},\ \bibinfo {pages} {235128} (\bibinfo {year}
  {2011})%
  \bibAnnoteFile{NoStop}{XieSPT2}%
\bibitem{XieSPT3}%
  \BibitemOpen
  \bibfield{author}{%
  \bibinfo {author} {\bibfnamefont{X.}~\bibnamefont{Chen}}, \bibinfo {author}
  {\bibfnamefont{Z.-X.}\ \bibnamefont{Liu}},\ and\ \bibinfo {author}
  {\bibfnamefont{X.-G.}\ \bibnamefont{Wen}},\ }%
  \bibfield{journal}{%
  \Doi{10.1103/PhysRevB.84.235141}{\bibinfo {journal} {Phys. Rev. B}}\ }%
  \textbf{\bibinfo {volume} {84}},\ \bibinfo {pages} {235141} (\bibinfo {year}
  {2011})%
  \bibAnnoteFile{NoStop}{XieSPT3}%
\bibitem{XieSPT4}%
  \BibitemOpen
  \bibfield{author}{%
  \bibinfo {author} {\bibfnamefont{X.}~\bibnamefont{Chen}}, \bibinfo {author}
  {\bibfnamefont{Z.-C.}\ \bibnamefont{Gu}}, \bibinfo {author}
  {\bibfnamefont{Z.-X.}\ \bibnamefont{Liu}},\ and\ \bibinfo {author}
  {\bibfnamefont{X.-G.}\ \bibnamefont{Wen}},\ \bibinfo {pages}
  {arXiv:1106.4772~}}%
   (\bibinfo {year} {2011})%
  \bibAnnoteFile{NoStop}{XieSPT4}%
\bibitem{PollmannSPT1}%
  \BibitemOpen
  \bibfield{author}{%
  \bibinfo {author} {\bibfnamefont{F.}~\bibnamefont{Pollmann}}, \bibinfo
  {author} {\bibfnamefont{E.}~\bibnamefont{Berg}}, \bibinfo {author}
  {\bibfnamefont{A.~M.}\ \bibnamefont{Turner}},\ and\ \bibinfo {author}
  {\bibfnamefont{M.}~\bibnamefont{Oshikawa}},\ \bibinfo {pages}
  {arXiv:0909.4059~}}%
   (\bibinfo {year} {2009})%
  \bibAnnoteFile{NoStop}{PollmannSPT1}%
\bibitem{PollmannSPT2}%
  \BibitemOpen
  \bibfield{author}{%
  \bibinfo {author} {\bibfnamefont{F.}~\bibnamefont{Pollmann}}, \bibinfo
  {author} {\bibfnamefont{A.~M.}\ \bibnamefont{Tuner}}, \bibinfo {author}
  {\bibfnamefont{e.}~\bibnamefont{Berg}},\ and\ \bibinfo {author}
  {\bibfnamefont{M.}~\bibnamefont{Oshikawa}},\ }%
  \bibfield{journal}{%
  \Doi{10.1103/PhysRevB.81.064439}{\bibinfo {journal} {Phys. Rev. B}}\ }%
  \textbf{\bibinfo {volume} {81}},\ \bibinfo {pages} {064439} (\bibinfo {year}
  {2010})%
  \bibAnnoteFile{NoStop}{PollmannSPT2}%
\bibitem{NorbertSPT}%
  \BibitemOpen
  \bibfield{author}{%
  \bibinfo {author} {\bibfnamefont{N.}~\bibnamefont{Schuch}}, \bibinfo {author}
  {\bibfnamefont{D.}~\bibnamefont{Perez-Garcia}},\ and\ \bibinfo {author}
  {\bibfnamefont{I.}~\bibnamefont{Cirac}},\ }%
  \bibfield{journal}{%
  \Doi{10.1103/PhysRevB.84.165139}{\bibinfo {journal} {Phys. Rev. B}}\ }%
  \textbf{\bibinfo {volume} {84}},\ \bibinfo {pages} {165139} (\bibinfo {year}
  {2011})%
  \bibAnnoteFile{NoStop}{NorbertSPT}%
\bibitem{LukaszfSPT1}%
  \BibitemOpen
  \bibfield{author}{%
  \bibinfo {author} {\bibfnamefont{L.}~\bibnamefont{Fidkowski}}\ and\ \bibinfo
  {author} {\bibfnamefont{A.}~\bibnamefont{Kitaev}},\ }%
  \bibfield{journal}{%
  \Doi{10.1103/PhysRevB.83.075103}{\bibinfo {journal} {Phys. Rev. B}}\ }%
  \textbf{\bibinfo {volume} {83}},\ \bibinfo {pages} {075103} (\bibinfo {year}
  {2011})%
  \bibAnnoteFile{NoStop}{LukaszfSPT1}%
\bibitem{Haldanephase}%
  \BibitemOpen
  \bibfield{author}{%
  \bibinfo {author} {\bibfnamefont{F.}~\bibnamefont{Haldane}},\ }%
  \bibfield{journal}{%
  \bibinfo {journal} {Phys. Lett. A}\ }%
  \textbf{\bibinfo {volume} {93}},\ \bibinfo {pages} {464} (\bibinfo {year}
  {1983})%
  \bibAnnoteFile{NoStop}{Haldanephase}%
\bibitem{RyuSPT}%
  \BibitemOpen
  \bibfield{author}{%
  \bibinfo {author} {\bibfnamefont{A.}~\bibnamefont{Schnyder}}, \bibinfo
  {author} {\bibfnamefont{S.}~\bibnamefont{Ryu}}, \bibinfo {author}
  {\bibfnamefont{A.}~\bibnamefont{Furusaki}},\ and\ \bibinfo {author}
  {\bibfnamefont{A.~W.~W.}\ \bibnamefont{Ludwig}},\ }%
  \bibfield{journal}{%
  \Doi{10.1103/PhysRevB.78.195125}{\bibinfo {journal} {Phys. Rev. B}}\ }%
  \textbf{\bibinfo {volume} {78}},\ \bibinfo {pages} {195125} (\bibinfo {year}
  {2008})%
  \bibAnnoteFile{NoStop}{RyuSPT}%
\bibitem{Kitaevperiod}%
  \BibitemOpen
  \bibfield{author}{%
  \bibinfo {author} {\bibfnamefont{A.~Y.}\ \bibnamefont{Kitaev}},\ }%
  \bibfield{journal}{%
  \bibinfo {journal} {AIP Conf. Proc.}\ }%
  \textbf{\bibinfo {volume} {1134}},\ \bibinfo {pages} {22} (\bibinfo {year}
  {2009})%
  \bibAnnoteFile{NoStop}{Kitaevperiod}%
\bibitem{FuKanepump}%
  \BibitemOpen
  \bibfield{author}{%
  \bibinfo {author} {\bibfnamefont{L.}~\bibnamefont{Fu}}\ and\ \bibinfo
  {author} {\bibfnamefont{C.~L.}\ \bibnamefont{Kane}},\ }%
  \bibfield{journal}{%
  \Doi{10.1103/PhysRevB.76.195312}{\bibinfo {journal} {Phys. Rev. B}}\ }%
  \textbf{\bibinfo {volume} {74}},\ \bibinfo {pages} {195312} (\bibinfo {year}
  {2006})%
  \bibAnnoteFile{NoStop}{FuKanepump}%
\bibitem{FuKaneHall}%
  \BibitemOpen
  \bibfield{author}{%
  \bibinfo {author} {\bibfnamefont{L.}~\bibnamefont{Fu}}\ and\ \bibinfo
  {author} {\bibfnamefont{C.~L.}\ \bibnamefont{Kane}},\ }%
  \bibfield{journal}{%
  \Doi{10.1103/PhysRevB.76.045302}{\bibinfo {journal} {Phys. Rev. B}}\ }%
  \textbf{\bibinfo {volume} {76}},\ \bibinfo {pages} {045302} (\bibinfo {year}
  {2007})%
  \bibAnnoteFile{NoStop}{FuKaneHall}%
\bibitem{LevinBurnellKoch}%
  \BibitemOpen
  \bibfield{author}{%
  \bibinfo {author} {\bibfnamefont{M.}~\bibnamefont{Levin}}, \bibinfo {author}
  {\bibfnamefont{F.~J.}\ \bibnamefont{Burnell}}, \bibinfo {author}
  {\bibfnamefont{M.}~\bibnamefont{Koch-Janusz}},\ and\ \bibinfo {author}
  {\bibfnamefont{A.}~\bibnamefont{Stern}},\ }%
  \bibfield{journal}{%
  \Doi{10.1103/PhysRevB.84.235145}{\bibinfo {journal} {Phys. Rev. B}}\ }%
  \textbf{\bibinfo {volume} {84}},\ \bibinfo {pages} {235145} (\bibinfo {year}
  {2011})%
  \bibAnnoteFile{NoStop}{LevinBurnellKoch}%
\bibitem{LukaszfSPT2}%
  \BibitemOpen
  \bibfield{author}{%
  \bibinfo {author} {\bibfnamefont{L.}~\bibnamefont{Fidkowski}}\ and\ \bibinfo
  {author} {\bibfnamefont{A.}~\bibnamefont{Kitaev}},\ }%
  \bibfield{journal}{%
  \Doi{10.1103/PhysRevB.81.134509}{\bibinfo {journal} {Phys. Rev. B}}\ }%
  \textbf{\bibinfo {volume} {81}},\ \bibinfo {pages} {134509} (\bibinfo {year}
  {2010})%
  \bibAnnoteFile{NoStop}{LukaszfSPT2}%
\bibitem{Wittencohomology}%
  \BibitemOpen
  \bibfield{author}{%
  \bibinfo {author} {\bibfnamefont{R.}~\bibnamefont{Dijkgraaf}}\ and\ \bibinfo
  {author} {\bibfnamefont{E.}~\bibnamefont{Witten}},\ }%
  \bibfield{journal}{%
  \Doi{10.1007/BF02096988}{\bibinfo {journal} {Commun. Math. Phys.}}\ }%
  \textbf{\bibinfo {volume} {129}},\ \bibinfo {pages} {393} (\bibinfo {year}
  {1990})%
  \bibAnnoteFile{NoStop}{Wittencohomology}%
\bibitem{Kogut}%
  \BibitemOpen
  \bibfield{author}{%
  \bibinfo {author} {\bibfnamefont{J.~B.}\ \bibnamefont{Kogut}},\ }%
  \bibfield{journal}{%
  \Doi{10.1103/RevModPhys.51.659}{\bibinfo {journal} {Rev. Mod. Phys.}}\ }%
  \textbf{\bibinfo {volume} {51}},\ \bibinfo {pages} {659} (\bibinfo {year}
  {1979})%
  \bibAnnoteFile{NoStop}{Kogut}%
\bibitem{KitaevToric}%
  \BibitemOpen
  \bibfield{author}{%
  \bibinfo {author} {\bibfnamefont{A.~Y.}\ \bibnamefont{Kitaev}},\ }%
  \bibfield{journal}{%
  \bibinfo {journal} {Annals of Physics}\ }%
  \textbf{\bibinfo {volume} {303}},\ \bibinfo {pages} {2} (\bibinfo {year}
  {2003})%
  \bibAnnoteFile{NoStop}{KitaevToric}%
\bibitem{Stringnet}%
  \BibitemOpen
  \bibfield{author}{%
  \bibinfo {author} {\bibfnamefont{M.~A.}\ \bibnamefont{Levin}}\ and\ \bibinfo
  {author} {\bibfnamefont{X.-G.}\ \bibnamefont{Wen}},\ }%
  \bibfield{journal}{%
  \Doi{10.1103/PhysRevB.71.045110}{\bibinfo {journal} {Phys. Rev. B}}\ }%
  \textbf{\bibinfo {volume} {71}},\ \bibinfo {pages} {045110} (\bibinfo {year}
  {2005})%
  \bibAnnoteFile{NoStop}{Stringnet}%
\bibitem{LevinWenHop}%
  \BibitemOpen
  \bibfield{author}{%
  \bibinfo {author} {\bibfnamefont{M.}~\bibnamefont{Levin}}\ and\ \bibinfo
  {author} {\bibfnamefont{X.-G.}\ \bibnamefont{Wen}},\ }%
  \bibfield{journal}{%
  \Doi{10.1103/PhysRevB.67.245316}{\bibinfo {journal} {Phys. Rev. B}}\ }%
  \textbf{\bibinfo {volume} {67}},\ \bibinfo {pages} {245316} (\bibinfo {year}
  {2003})%
  \bibAnnoteFile{NoStop}{LevinWenHop}%
\bibitem{Note1}%
  \BibitemOpen
  \bibinfo {note} {To be precise, the $\sigma ^x_p O_p$ terms at the sites
  neighboring the endpoints of $\beta $ do not commute with $V^0_\beta $, but
  the excitations associated with these terms should be regarded as part of the
  $\pi $-flux excitation.}%
  \bibAnnoteFile{Stop}{Note1}%
\bibitem{Wegner}%
  \BibitemOpen
  \bibfield{author}{%
  \bibinfo {author} {\bibfnamefont{F.}~\bibnamefont{Wegner}},\ }%
  \bibfield{journal}{%
  \bibinfo {journal} {J. Math. Phys.}\ }%
  \textbf{\bibinfo {volume} {12}},\ \bibinfo {pages} {2259} (\bibinfo {year}
  {1971})%
  \bibAnnoteFile{NoStop}{Wegner}%
\bibitem{Kitaevhoneycomb}%
  \BibitemOpen
  \bibfield{author}{%
  \bibinfo {author} {\bibfnamefont{A.}~\bibnamefont{Kitaev}},\ }%
  \bibfield{journal}{%
  \Doi{10.1016/j.aop.2005.10.005}{\bibinfo {journal} {Ann. Phys.}}\ }%
  \textbf{\bibinfo {volume} {321}},\ \bibinfo {pages} {2} (\bibinfo {year}
  {2006})%
  \bibAnnoteFile{NoStop}{Kitaevhoneycomb}%
\bibitem{Note2}%
  \BibitemOpen
  \bibinfo {note} {For readers familiar with the string-net models in Ref.
  \protect \rev@citealpnum {Stringnet}, we note that one can equivalently
  construct string-net models by letting the string types be \protect \emph
  {irreducible representations} of $G$, but these models are not as convenient
  for the duality construction when $G$ is non-abelian.}%
  \bibAnnoteFile{Stop}{Note2}%
\bibitem{Note3}%
  \BibitemOpen
  \bibinfo {note} {This relation -- which is closely related to the path
  independence of the string operator $V_\beta ^1$ -- follows from the operator
  identity $W_\beta = W_{\beta '} \DOTSB \prod@ \slimits@ _{p \in \beta '
  -\beta } B_p$, where the product runs over sites $p$ in between the two paths
  $\beta , \beta '$.}%
  \bibAnnoteFile{Stop}{Note3}%
\bibitem{Hastingsadiab}%
  \BibitemOpen
  \bibfield{author}{%
  \bibinfo {author} {\bibfnamefont{M.~B.}\ \bibnamefont{Hastings}}\ and\
  \bibinfo {author} {\bibfnamefont{X.-G.}\ \bibnamefont{Wen}},\ }%
  \bibfield{journal}{%
  \Doi{10.1103/PhysRevB.72.045141}{\bibinfo {journal} {Phys. Rev. B}}\ }%
  \textbf{\bibinfo {volume} {72}},\ \bibinfo {pages} {045141} (\bibinfo {year}
  {2005})%
  \bibAnnoteFile{NoStop}{Hastingsadiab}%
\bibitem{SubirBook}%
  \BibitemOpen
  \bibfield{author}{%
  \bibinfo {author} {\bibfnamefont{S.}~\bibnamefont{Sachdev}},\ }%
  \emph{\bibinfo {title} {Quantum phase transitions}}\ (\bibinfo {publisher}
  {Cambridge University Press},\ \bibinfo {year} {1999})%
  \bibAnnoteFile{NoStop}{SubirBook}%
\bibitem{WenBook}%
  \BibitemOpen
  \bibfield{author}{%
  \bibinfo {author} {\bibfnamefont{X.-G.}\ \bibnamefont{Wen}},\ }%
  \emph{\bibinfo {title} {Quantum Field Theory of Many-body Systems: From the
  Origin of Sound to an Origin of Light and Electrons}}\ (\bibinfo {publisher}
  {Oxford University Press},\ \bibinfo {year} {2007})%
  \bibAnnoteFile{NoStop}{WenBook}%
\bibitem{TV}%
  \BibitemOpen
  \bibfield{author}{%
  \bibinfo {author} {\bibfnamefont{V.~G.}\ \bibnamefont{Turaev}}\ and\ \bibinfo
  {author} {\bibfnamefont{O.~Y.}\ \bibnamefont{Viro}},\ }%
  \bibfield{journal}{%
  \bibinfo {journal} {Topology}\ }%
  \textbf{\bibinfo {volume} {31}},\ \bibinfo {pages} {865} (\bibinfo {year}
  {1992})%
  \bibAnnoteFile{NoStop}{TV}%
\end{thebibliography}%

\end{document}